\documentclass[12pt]{article}
\usepackage{url, natbib}
\usepackage{graphicx}
\usepackage{amsmath, color, adjustbox, bm}

\renewcommand{\baselinestretch}{1.2}

\author{ \hspace{-1cm}\small
Yun-Hee Choi$^{1,*}$, Hae Jung$^{1}$, Saundra Buys$^{3}$, Mary Daly$^{4}$, Esther M. John$^{5}$, \\
\small  \hspace{-1cm}
John Hopper$^{6}$, Irene Andrulis$^{2}$, Mary Beth Terry$^{7}$, Laurent Briollais$^{2,8}$ \\ \\
\small \hspace{-1cm}
$^{1,\ast}$Department of Epidemiology and Biostatistics, University of Western Ontario, London, Canada \\
\small \hspace{-2cm}
$^2$ Lunenfeld-Tanenbaum Research Institute, Mount Sinai Hospital, Toronto, Canada \\
\small\hspace{-2cm}
$^3$ University of Utah Health Sciences Center, Salt Lake City, Utah, USA \\
\small \hspace{-2cm}
$^4$ Fox Chase Cancer Center, Philadelphia, Pennsylvania, USA \\
\small \hspace{-2cm}
$^5$ Stanford University School of Medicine, Stanford, California, USA \\
\small \hspace{-2cm}
$^6$ School of Population and Global Health, The University of Melbourne, Carlton, Victoria, Australia \\
\small \hspace{-2cm}
$^7$ Mailman School of Public Health, Columbia University, New York, USA \\
\small \hspace{-2cm}
$^8$ Division of Biostatistics, Dalla Lana School of Public Health, University of Toronto, Canada \\ \\
{\small \hspace{-2cm} $^{*}$ Yun-Hee.Choi@schulich.uwo.ca}
}
\title {A Competing Risks Model with \\
Binary Time Varying Covariates for 
Estimation of \\ Breast Cancer Risks in \textit{BRCA1} Families}
\date{}
\begin{document}
\maketitle

\begin{abstract}
{Mammographic screening and prophylactic surgery such as risk-reducing salpingo oophorectomy (RRSO) can potentially reduce breast cancer risks among mutation carriers of \textit{BRCA} families. The evaluation of these interventions is usually complicated by the fact that their effects on breast cancer may change over time and by the presence of competing risks. We introduce a correlated competing risks model to model breast and ovarian cancer risks within \textit{BRCA1} families that accounts for time-varying covariates (TVCs). Different parametric forms for these TVCs are proposed for more flexibility and a correlated gamma frailty model is specified to account for the correlated competing events. We also introduced a new ascertainment correction approach that accounts for the selection of families  through probands affected with either breast or ovarian cancer, or unaffected. Our simulation studies demonstrate the good performances of our proposed approach in terms of bias and precision of the estimators of model parameters  and cause-specific penetrances over different levels of familial correlations. We apply our new approach to 498 \textit{BRCA1} mutation carrier families recruited through the Breast Cancer Family Registry. Our results demonstrate the importance of the functional form of the TVC when assessing the role of RRSO on breast cancer. In particular, under the best fitting TVC model, the overall effect of RRSO on breast cancer risk was statistically significant in women with \textit{BRCA1}.
}
{Breast and ovarian cancers; \textit{BRCA}; Competing risks; Time-varying covariate; Correlated frailty model; Penetrance; Risk-reducing salpingo oophorectomy.}
\end{abstract}

\section{Introduction}
\label{s:intro}
Between 10-15\% of all breast cancers (BCs) are caused by a hereditary predisposition \cite{Aloraifi15}. Hereditary Breast and Ovarian Cancer syndrome (HBOC) is an autosomal dominant disease characterized by germline pathogenic mutations in the {\it BRCA1} and {\it BRCA2} genes for the majority of cases. It is the most common cause of hereditary forms of both breast and ovarian cancer (OC) \cite{Petrucelli10}. The overall prevalence of {\it BRCA1/2} mutations is estimated to be from 1 in 400 to 1 in 800 with a higher prevalence in the Ashkenazi Jewish population (1 in 40). Estimates of penetrance (cancer risk) for {\it BRCA1/2} mutations vary considerably \cite{Petrucelli10}. Previous large meta-analyses reported mean cumulative BC risks at age 70 of 57\% for {\it BRCA1} and 49\% for {\it BRCA2} mutation carriers \cite{Chen07, Kuchenbaecker17}. The OC risks were 40\% for {\it BRCA1} and 18\% for {\it BRCA2} mutation carriers. Mutation carriers are also at an elevated risk of developing contralateral breast cancer (CBC) after a previous unilateral BC \cite{Kuchenbaecker17}. A recent meta-analysis estimated the 5-year CBC risk at 15\% for {\it BRCA1} mutation carriers and 9\% for {\it BRCA2} mutation carriers after a first BC \cite{Molina-Montes14}. Risk prediction models can be used to assess these risks in {\it BRCA1/2} mutation positive families. These statistical models can help health practitioners to guide women who could benefit from genetic counselling and also in their clinical management, which currently comprise intensified surveillance for early BC detection using multimodal imaging techniques or prophylactic surgery such as bilateral mastectomy for the risk of BC and risk-reducing salpingo-oophorectomy (RRSO) for the risk of OC \cite{Practice17}.

Competing risks models for clustered failure times data have already been proposed by Gorfine and Hsu\cite{Gorfine2011}, which extended the competing risks model of Prentice et al \cite{Prentice78} to incorporate the frailty variables to cause-specific hazards models for all the causes. In a subsequent paper, Gorfine et al \cite{Gorfine2014} showed through a simulation study that naively treating competing risks as independent right censoring events resulted in non-calibrated predictions of cancer risks, with the expected number of events overestimated. Recently, we have also proposed a competing risks approach for clustered family data applicable to successive time-to-event outcomes (i.e. the first and second cancer event could each have a competing risk event) \cite{Choi17}. However, to our knowledge, none of these approaches was developed to include time varying covariates (TVCs). 

In clinical setting, assessing the effect of TVCs is important especially when the follow-up duration is long. For example, we can consider a binary variable for a certain treatment occurring at a later period of the follow-up duration. If we code this variable as time invariant covariate (TIC), the duration of treatment exposure becomes much longer than the actual exposure. We lose the information that the subject was actually absent of its effect for most part of the follow-up period. This type of TVC is referred to as permanent exposure (PE) as its effect stays constant permanently since the time of the treatment exposure.  The formulation of TVC effect, which decays over time with the rate parameter, is referred to as exponential decay (ED) \cite{Keown-Stoneman18}. Cox and Oakes \cite{Cox84} include an additional parameter that measures the converged effect of TVC, referred to as Cox and Oakes (CO) model.

In this paper, our goal is to extend previous competing risks approach \cite{Choi17, Gorfine2011} to the situation where the cause-specific hazard function for the main event of interest, BC, can depend on TVCs such as mammography screening (MS) or RRSO. The second main extension is to propose an ascertainment correction that specifically accounts for the fact that the {\it BRCA1} families have been recruited through a proband affected by either BC or OC before her study entry, or through an unaffected proband. With our proposed approach, we have BC, OC and death from other causes as competing events in {\it BRCA1} mutation families.  We also demonstrated a very relevant application of our model to a large series of {\it BRCA1} families, in particular, with an assessment of RRSO. The possibility that RRSO prevents future BC has been the subject of some debate. Terry et al \cite{Terry19b} did not find an association after accounting for the time-varying nature of the covariate. There may be some benefit in RRSO, however women may elect for RRSO close to menopause limiting the impact. Here we consider the impact of the timing of RRSO in addition to MS through both simulations and applied analyses.  

\section{Methods}
\label{s:model}

\subsection{Correlated gamma frailty model for competing events with time-varying covariates}

Consider data arising from $n$ independent families, with family $f$, $f=1,\ldots,n$, each family consisting of $n_f$ members, $i = 1, \ldots, n_f$. For family member $i$ in family $f$, we denote by $T^\ast_{f_{i}}$ and $C_{f_i}$ the time to the first event time and the right censoring time, respectively, and by $\delta_{f_{i}} \in \{1,\ldots,J\}$ the type of the first observed event among $J$ competing events and $\delta_{f_{i}}=0$ if right censored. The observed time is then defined as $T_{f_{i}}=\mathrm{min}(T^\ast_{f_{i}},C_{f_i})$. We denote by $Z_{f_j}$ the unobserved frailty shared within family $f$ for event $j \;\; (j=1, \cdots, J)$. To allow covariates to vary over time, let  $\bm{x}_{f_{i}}(t)$ be the vector of TVCs at time $t$ for individual $i$ in family $f$ and $\bm{X}_{f_i}(t) = \{\bm{x}_{f_i}(u); 0 \le u < t  \}$ represent the covariate history up to time $t$. Then the cause-specific hazard function for event $j$ for individual $i$ from family $f$ conditional on the covariate history $\bm{X}_{f_{i}}(t)$ and cause-specific familial frailty $Z_{f_j}$ follows a proportional hazards regression model
\begin{eqnarray}
h_{f_{ij}}(t|\bm{X}_{f_i}(t),Z_{f_j})
&=&\lim_{dt\to 0}\frac{1}{dt}P(t \leq T^\ast_{f_{i}} < t+dt, \delta_{f_{i}}=j|T^\ast_{f_{i}} \geq t,\bm{X}_{f_i}(t),Z_{f_j}) \nonumber \\
&=&h_{0j}(t) Z_{f_j} e^{\bm{\beta}_j^T\bm{x}_{f_i}(t) },\label{eq:1}
\end{eqnarray}
where $h_{0j}(t)$ is the baseline hazard function and $\bm{\beta}_j$ is the vector of the covariate effects related to event $j$. We assume the time-varying covariates are exogenous---the future values of covariates up to any time $t > u$ are not affected by the occurrence of any event at time $u$.  

The family-specific frailties $Z_{f_j}$ for event $j$ are random effects shared within families. We assume that the frailties are independent across families given event $j$, but the event-specific frailties could be correlated with each other within families. The correlated frailties can be constructed by defining 
 each event-specific frailty $Z_{f_j}$ within families using two independent random variables  $Y_{f_0}$ and $Y_{f_j}$ \cite{Yashin95, Wienke11} so that any pair of family members with different events shares the common frailty $Y_{f_0}$ to induce possible dependence across competing events within families. Gamma frailties are commonly used in the literature because of their mathematical convenience for constructing likelihoods with close-form expression. Other distributions such as log-normal or compound Poisson distributions can be used as well for frailties. For correlated log-normal frailties, a multivariate log-normal distribution can be directly used to construct the dependence via the covariance matrix. However, there is no close form expression for such distribution when integrating out the frailties to construct marginal likelihood and numerical integration is needed. In our paper, we present correlated gamma frailties to provide close form expressions of marginal likelihood and cause-specific penetrance functions, i.e., absolute risk of event given the mutation status for each individual.

We construct the correlated gamma frailties by defining $$Z_{f_j} = \frac{\omega_0}{\omega_j} Y_{f_0} + Y_{f_j},$$  
where $Y_{f_0}, Y_{f_j}, j=1, \ldots, J$ are independent gamma distributed frailties following $Y_{f_0} \sim \mbox{Gamma}(k_0, 1/k_0)$ and $Y_{f_j} \sim \mbox{Gamma}(k_j, 1/(k_0+k_j))$ and $\omega_0=k_0, \omega_j = k_0 + k_j.$ Then, 
$Z_{f_j}$ follows $\mbox{Gamma}(\omega_j, 1/\omega_j )$ with mean 1 and variance $ = 1/\omega_j $ and the covariance of the frailties of two events $j$ and $j'$, $j\ne j'$, can be expressed as $\mbox{cov}(Z_{f_j}, Z_{f_{j'}}) = \frac{\omega_0}{   \omega_j \omega_{j'} }$, and the correlation as $\rho = \frac{\omega_0}{\sqrt{\omega_j \omega_{j'}}}$.
As a special case, $\omega_0=0$ corresponds to the independent frailties.

The overall survival function is defined as the probability of surviving from all competing events conditional on the covariate history and frailties:
\begin{equation}
S_{f_{i}}(t|\bm{X}_{f_{i}}(t), \bm{Z}_f) = \mathrm{exp}\bigg\{-\sum_{j=1}^{J} H_{f_{ij}}(t|\bm{X}_{f_{i}}(t),Z_{f_j})\bigg\} \label{eq:3} ,
\end{equation}
where $\bm{Z}_f = \{ Z_{f_1}, \ldots, Z_{f_J} \} $ and
$H_{f_{ij}}(t|\bm{X}_{f_{i}}(t),Z_{f_j}) = \int_{0}^{t} h_{0j}(u) Z_{f_j} e^{\bm{\beta}_j^T\bm{x}_{f_i}(u) } du$ is the cause-specific cumulative hazard function at time $t$. \\

Consider a binary time varying covariate $x_{f_i}(t)$ = 0 at $t<t_x$ and 1 at $t \geq t_x$, where $t_x$ is the time that changes in value of covariate occurred. We can describe  the effect of the TVC that changes over time, denoted by $\mu(\cdot)$,  in three different structures: PE, ED, and CO as follows,
$$
\mu(x_{f_i}(t))=
\begin{cases} 
   0                                                       & \quad \text{ if }  t < t_x  \; \mathrm{  (PE,ED,CO) }\\
    \beta                                                       & \quad \text{ if } t \geq t_x \text{ (PE) }\\
    \beta\, \mathrm{exp}\big\{-\eta(t-t_x)\big\}                     & \quad \text{ if } t \geq t_x \text{ (ED) }\\
    \beta\, \mathrm{exp}\big\{-\eta(t-t_x)\big\} + \eta_0          & \quad \text{ if } t \geq t_x \text{ (CO) },
\end{cases}
$$
where for time $t \ge t_x$, the effect of TVC stays at $\beta$ for PE, whereas it starts to decrease exponentially with a rate of $e^{-\eta}$ to 0 for ED or to $\eta_0$ for CO. The $j$th cause-specific hazard and cumulative hazard function with TVC can be written as 
$$h_{f_{ij}}(t| X_{f_i}(t), Z_{f_j})= h_{0j}(t)Z_{f_j}\mathrm{exp}\big\{ \mu(x_{f_i}(t))  \big\}, $$
$$H_{f_{ij}}(t|X_{f_i}(t), Z_{f_j})= \int_0^{t} h_{0j}(u)Z_{f_j}\mathrm{exp}\big\{ \mu(x_{f_i}(u))  \big\} du, $$
where calculation details for cause-specific cumulative hazard for PE, ED and CO models are specified in Web Appendix A.

\subsection{Likelihood construction}

Let $\bm{\theta}=\{h_{0j}(.), \bm{\beta}_j, k_0, k_j, \eta_j, \eta_{0j}, j=1,\ldots, J\}$ be the vector of parameters involved in the the model, which consists of baseline parameters for specifying baseline hazard functions, regression coefficient vector $\bm{\beta}_j$, $\eta_j$ and $\eta_{0j}$, related to TVC effects, and frailty parameters $k_0, k_j$ for competing event $j=1,...,J$. Then, the likelihood of the data from $n$ families can be constructed simply by the product of the likelihoods of all families: $$L(\bm{\theta}) = \prod_{f=1}^{n} L_f(\bm{\theta}).$$

Under the shared frailty competing risk model framework, the likelihood for family $f$ is obtained by integrating over the frailty distribution:
\begin{eqnarray*}
L_f(\bm{\theta}) &=& 
\prod_{i=1}^{n_f} \int_{0}^{\infty} \cdots \int_{0}^{\infty}  \left\{ \prod_{j=1}^{J}  h_{f_{ij}}(t_{f_{i}}|\bm{X}_{f_{i}}(t_{f_i}),Z_{f_j})^{\mathrm{I}(\delta_{f_i}=j)}  \right \} \times \\
& &S_{f_{i}}(t_{f_{i}}|\bm{X}_{f_{i}}(t_{f_i}),\bm{Z}_{f}) g_Z(Z_{f_1}, \ldots, Z_{f_J})dZ_{f_1}\dots dZ_{f_J}.
\end{eqnarray*}

To compute the integrals, we replace $Z_{f_j}$ by $Y_{f_0} + Y_{f_j}, j=1, \ldots, J$ and integrate out the independent random variables $Y_{f_m}, m=0, \ldots, J$,  utilizing their Laplace transform $\phi_m(\cdot)$ and their $d$th derivative, $\phi_m(\cdot)^{(d)}$, which have the following expressions
\begin{eqnarray*}
\phi_m(s)&=&\int_{0}^{\infty} e^{-sz}g_m(z)dz \\
\phi_m(s)^{(d)}&=&(-1)^{d} \int_{0}^{\infty} z^d e^{-sz}g_m(z)dz, 
\end{eqnarray*}
where $g_m(\cdot)$ represents the density function of the random variable $Y_{f_m}$. 

With  $Y_{f_m} \sim \text{Gamma}(k_m,\frac{1}{\omega_m})$, $\omega_0 = k_0, \omega_m = k_0 + k_m, m\ne 0,$
they have closed form expressions:
\begin{eqnarray*}
\phi_m(s)&=& \left( 1+ \frac{s}{\omega_m}\right)^{-k_m} \\
\phi_m(s)^{(d)}&=&(-1)^{d} \frac{\Gamma(k_m+d)}{\Gamma(k_m) \, w_m^{d}}\left( 1+ \frac{s}{\omega_m}\right)^{-k_m-d}.
\end{eqnarray*}

Thus, the likelihood for family $f$ can be obtained as
\begin{eqnarray}
L_f(\bm{\theta}) &=& 
\prod_{i=1}^{n_f} \int_{0}^{\infty} \int_{0}^{\infty} \cdots \int_{0}^{\infty}  \prod_{j=1}^{J}  h_{f_{ij}}(t_{f_{i}}|\bm{X}_{f_{i}}(t_{f_i}), Y_{f_0}, Y_{f_j})^{\mathrm{I}(\delta_{f_i}=j)}   \times \\ \nonumber
& &S_{f_{i}}(t_{f_{i}}|\bm{X}_{f_{i}}(t_{f_i}),\bm{Y}_{f}) g_0(Y_{f_0}) g_1(Y_{f_1}), \ldots, g_J(Y_{f_J})dY_{f_0} dY_{f_1}\dots dY_{f_J} \\ \nonumber
&=& \prod_{i=1}^{n_f}  \int_{0}^{\infty} \int_{0}^{\infty} \cdots \int_{0}^{\infty}    \prod_{j=1}^{J} 
           \left \{ \left (\frac{\omega_0}{\omega_j}Y_{f_0} + Y_{f_j} \right ) h_{ij}(t_{f_i}|\bm{X}_{f_i}(t_{f_i}))  \right \}^{\mathrm{I}(\delta_{f_i}=j)}  \times \\ \nonumber
&  &   e^{-\sum_{j=1}^J \left (\frac{\omega_0}{\omega_j}Y_{f_0} + Y_{f_j} \right ) \sum_{i=1}^{n_f}  H_{ij}(t_{f_i}|\bm{X}_{f_i}(t_{f_i}))} 
g_0(Y_{f_0})g_1(Y_{f_1}), \ldots, g_J(Y_{f_J})dY_{f_0} dY_{f_1}\dots dY_{f_J}  \\ \nonumber
&=& \left\{ \prod_{i=1}^{n_f}  \prod_{j=1}^J  h_{ij}(t_{f_i}|\bm{X}_{f_i}(t_{f_i}))^{\mathrm{I}(\delta_{f_i}=j)}  \right \}
 \int_{0}^{\infty} \int_{0}^{\infty} \cdots \int_{0}^{\infty}     
             \prod_{j=1}^{J}  \left (\frac{\omega_0}{\omega_j} Y_{f_0} + Y_{f_j} \right )^{ d_{f_j} } \times \\ \nonumber
&  &   e^{- Y_{f_0} \left  \{ \sum_{j=1}^J \frac{\omega_0}{\omega_j}    \sum_{i=1}^{n_f} H_{ij}(t_{f_i}|\bm{X}_{f_i}(t_{f_i})) \right \}  
- \sum_{j=1}^J    Y_{f_j}  \left \{  \sum_{i=1}^{n_f} H_{ij}(t_{f_i}|\bm{X}_{f_i}(t_{f_i})) \right \}  } \times \\ \nonumber
& & g_0(Y_{f_0})g_1(Y_{f_1}), \ldots, g_J(Y_{f_J})dY_{f_0} dY_{f_1}\dots dY_{f_J}  \\ \nonumber
&=& \left\{ \prod_{i=1}^{n_f}  \prod_{j=1}^J  h_{ij}(t_{f_i}|\bm{X}_{f_i}(t_{f_i}))^{\mathrm{I}(\delta_{f_i}=j)}  \right \}
 \sum_{x_1=0}^{d_{f_1}} \cdots \sum_{x_J=0}^{d_{f_J}}  
  \int_{0}^{\infty} \int_{0}^{\infty} \cdots \int_{0}^{\infty} \\ \nonumber 
& & 
 Y_{f_0}^{\sum_{j=1}^J x_j}  
 e^{- Y_{f_0} \left  \{ \sum_{j=1}^J \frac{\omega_0}{\omega_j}    \dot{H}_j \right \} } 
     \left \{ \prod_{j=1}^J { d_{f_j} \choose x_j} \left ( \frac{\omega_0}{\omega_j}  \right)^{x_j}   
 Y_{f_j}^{d_{f_j} - x_j} \right \}
 e^{ - \sum_{j=1}^J    Y_{f_j}    \dot{H}_j   } \times \\ \nonumber
& & g_0(Y_{f_0})g_1(Y_{f_1}), \ldots, g_J(Y_{f_J})dY_{f_0} dY_{f_1}\dots dY_{f_J}  \\ \nonumber
&=& \left\{ \prod_{i=1}^{n_f}  \prod_{j=1}^J  h_{ij}(t_{f_i}|\bm{X}_{f_i}(t_{f_i}))^{\mathrm{I}(\delta_{f_i}=j)}  \right \}
 \sum_{x_1=0}^{d_{f_1}} \cdots \sum_{x_J=0}^{d_{f_J}}  
 (-1)^{ \sum_{j=1}^J x_j } \phi_0^{ (\sum_{j=1}^J x_j) } \left ( \sum_{j=1}^J \frac{\omega_0}{\omega_j} \dot{H}_j \right )
 \times \\ \nonumber 
& & 
 \left \{ \prod_{j=1}^J { d_{f_j} \choose x_j} \left ( \frac{\omega_0}{\omega_j}  \right)^{x_j} 
  (-1)^{ d_{f_j} -  x_j } \phi_j^{ ( d_{f_j} -  x_j) } \left ( \dot{H}_j \right )   \right \} 
\end{eqnarray}
where $d_{f_j} = \sum_{i=1}^{n_f} \mathrm{I}(\delta_{f_i}=j) $ is the number of family members affected by event $j$,   
 $ \dot{H}_j  =  \sum_{i=1}^{n_f} H_{ij}(t_{f_i}|\bm{X}_{f_i}(t_{f_i})) $ is used for notational simplicity and 
the products of binomials are written using summations based on the binomial theorem, 
\begin{eqnarray*} 
\prod_{j=1}^{J}  \left (\frac{\omega_0}{\omega_j} Y_{f_0} + Y_{f_j} \right )^{ d_{f_j} } 
&=& 
\sum_{x_1=0}^{d_{f_1}} \cdots \sum_{x_J=0}^{d_{f_J}} { d_{f_1} \choose x_1}
\left ( \frac{\omega_0}{\omega_1} Y_{f_0} \right)^{x_1} Y_{f_1}^{d_{f_1} - x_1}
\cdots { d_{f_J} \choose x_J} \left ( \frac{\omega_0}{\omega_J} Y_{f_0} \right)^{x_J} Y_{f_J}^{d_{f_J} - x_J}   \\ 
&=& \sum_{x_1=0}^{d_{f_1}} \cdots \sum_{x_J=0}^{d_{f_J}}  { d_{f_1} \choose x_1} \left ( \frac{\omega_0}{\omega_1}  \right)^{x_1} 
\cdots { d_{f_J} \choose x_J} \left ( \frac{\omega_0}{\omega_1}  \right)^{x_J}
Y_{f_0}^{\sum_{j=1}^J x_j}  Y_{f_1}^{d_{f_1} - x_1} \cdots Y_{f_J}^{d_{f_J} - x_J} \\
&=& \sum_{x_1=0}^{d_{f_1}} \cdots \sum_{x_J=0}^{d_{f_J}}  Y_{f_0}^{\sum_{j=1}^J x_j}
\left \{ \prod_{j=1}^J { d_{f_j} \choose x_j} \left ( \frac{\omega_0}{\omega_j}  \right)^{x_j} Y_{f_j}^{d_{f_j} - x_j} \right \}.  \\
\end{eqnarray*}

With the Laplace transform of the gamma frailties, the likelihood can be further simplified as
\begin{eqnarray}
L_f(\bm{\theta}) &=& 
\left\{ \prod_{i=1}^{n_f}  \prod_{j=1}^J  h_{ij}(t_{f_i}|\bm{X}_{f_i}(t_{f_i}))^{\mathrm{I}(\delta_{f_i}=j)}  \right \} \times  \nonumber \\
 && \sum_{x_1=0}^{d_{f_1}} \cdots \sum_{x_J=0}^{d_{f_J}}  
 \frac{\Gamma(k_0+\sum_{j=1}^J x_j)}{\Gamma(k_0) \, k_0^{\sum_{j=1}^J x_j}}\left( 1+  \sum_{j=1}^J \frac{ \dot{H}_j }{k_0+ k_j}\right)^{-k_0-\sum_{j=1}^J x_j}
 \times  \nonumber \\ 
& & 
 \left \{ \prod_{j=1}^J { d_{f_j} \choose x_j} \left ( \frac{\omega_0}{\omega_j}  \right)^{x_j} 
 \frac{\Gamma(k_j+d_{f_j} -  x_j)}{\Gamma(k_j) \, (k_0+k_j)^{d_{f_j} -  x_j}}\left( 1+ \frac{ \dot{H}_j }{k_0+k_j}\right)^{-k_j-d_{f_j} +  x_j}  
\right \}  \nonumber \\
&=& \left\{ \prod_{i=1}^{n_f}  \prod_{j=1}^J  h_{ij}(t_{f_i}|\bm{X}_{f_i}(t_{f_i}))^{\mathrm{I}(\delta_{f_i}=j)}  \right \} \prod_{j=1}^J (k_0 + k_j)^{-d_{f_j}}   \times  \nonumber \\
 && \sum_{x_1=0}^{d_{f_1}} \cdots \sum_{x_J=0}^{d_{f_J}}  
 \frac{\Gamma(k_0+\sum_{j=1}^J x_j)}{\Gamma(k_0)  }\left( 1+  \sum_{j=1}^J \frac{ \dot{H}_j }{k_0+ k_j}\right)^{-k_0-\sum_{j=1}^J x_j}
 \times  \nonumber \\ 
& & 
 \left \{ \prod_{j=1}^J { d_{f_j} \choose x_j}  
 \frac{\Gamma(k_j+d_{f_j} -  x_j)}{\Gamma(k_j)  }  \left( 1+ \frac{ \dot{H}_j }{k_0+k_j}\right)^{-k_j-d_{f_j} +  x_j}  
\right \}.  \label{eqn:Lf}
\end{eqnarray}

\subsection{Ascertainment correction}

It is common in familial cancer studies that families are ascertained via a proband (indexed as $p$) who is affected with cancer. A correction for ascertainment needs to be applied to get valid inference about the penetrance function and genetic relative risk and we have previously proposed and evaluated several approaches for this problem in the context of a single time to event outcome \cite{Choi08}. We generalize here the prospective likelihood approach of ascertainment correction  
that we introduced before, to the situation where the proband has at least one of the three competing events (BC, OC or death from other causes) before her age at examination ($a_{f_p}$). The reason we also consider death as an ascertainment event is that in our real application, a small number of probands were unaffected at study entry but died during the follow-up period. \\

The rationale of the prospective likelihood method of ascertainment correction is to weight the likelihood of each family $f$, $L_f(\bm{\theta})$, by the inverse probability of a proband being selected before her age at examination, assuming the proband could have been ascertained anytime within this interval.
We denote this probability $A_f(\bm{\theta})=P(T_{f_p} \leq a_{f_p} | \bm{X}_{f_p}(a_{f_p}))$, which can be derived as
\begin{eqnarray}
A_f(\bm{\theta})
                 &=& 1 - \idotsint \mathrm{exp}\bigg\{- \sum_{j=1}^{J} Z_{f_j} H_{f_{pj}}(a_{f_p}|\bm{X}_{f_{p}} (a_{f_p}) ) \bigg\} 
		                                                              g_Z(Z_{f_1}, \dots, Z_{f_J})dZ_{f_1}\dots dZ_{f_J} \nonumber \\  
                 &=& 1-  \bigg \{ 1 + \sum_{j=1}^{J}  \frac{ H_{f_{pj}}(a_{f_p}|\bm{X}_{f_{p}} (a_{f_p}) ) } { \omega_j } \bigg\}^{-k_0} 
                 \prod_{j=1}^J \bigg\{ 1+ \frac{H_{f_{pj}}(a_{f_p}|\bm{X}_{f_{p}} (a_{f_p}) ) }{\omega_j}\bigg\}^{-k_j}.      
                 \label{eqn:den}         
\end{eqnarray}

In our real data application, we also consider unaffected probands. The ascertainment correction for them is given by the probability of surviving all events
$$A_f(\bm{\theta}) = \bigg \{ 1 + \sum_{j=1}^{J}  \frac{ H_{f_{pj}}(a_{f_p}|\bm{X}_{f_{p}} (a_{f_p}) ) } { \omega_j } \bigg\}^{-k_0} 
                 \prod_{j=1}^J \bigg\{ 1+ \frac{H_{f_{pj}}(a_{f_p}|\bm{X}_{f_{p}} (a_{f_p}) ) }{\omega_j}\bigg\}^{-k_j} .  $$

Therefore, the ascertainment corrected likelihood for all the families is expressed as $$L_C(\bm{\theta}) = \prod_{f=1}^{n} \frac{L_f(\bm{\theta})}{A_f(\bm{\theta})}, $$
and maximum likelihood estimates of the parameters are obtained by maximizing the corresponding log-likelihood. 
 
\subsection{Cause-specific penetrance function with time-varying covariates}

Our main interest is to estimate the $j$th cause-specific cumulative incidence function $F_j(\cdot)$, also called cause-specific penetrance. We first express the conditional cause-specific penetrance given the random frailty variables $\bm{Z} = \{Z_1, \ldots, Z_J\}$ as
\begin{eqnarray}
F_j(t|\bm{X}_{f_{i}}(t), \bm{Z})  
&=& P(T_{f_{i}} \leq t, \delta_{f_{i}}=j|\bm{X}_{f_{i}}(t), \bm{Z} ) \nonumber \\
&=& \int_{0}^{t} h_{f_{ij}}(u|\bm{X}_{f_{i}}(u),Z_{j})  \mathrm{exp}\bigg \{- \sum_{j=1}^{J} H_{f_{ij}}(u|\bm{X}_{f_{i}}(u),Z_{j}) \bigg\} du. \nonumber
\end{eqnarray}

We derived the marginal cause-specific penetrance function for event $j$ by integrating over the frailties ${\bm Z}=\{Z_1, \ldots, Z_J\}$ as follows:
\begin{eqnarray}
F_j(t|\bm{X}_{f_{i}}(t)) 
&=& \int_{0}^{\infty} \cdots \int_{0}^{\infty} 
        \int_{0}^{t} h_{f_{ij}}(u|\bm{X}_{f_{i}}(u), Z_j) S_{f_{i}}(u|\bm{X}_{f_{i}}(u), {\bm Z})     g_Z({\bm Z}) du d{\bm Z}  \nonumber \\
&=& \int_{0}^{t} \int_{0}^{\infty} \cdots  \int_{0}^{\infty} h_{f_{ij}} (u) \left(\frac{\omega_0}{\omega_j}Y_0 + Y_j\right) 
e^{ -  \sum_{l=1}^J \left(\frac{\omega_0}{\omega_l}Y_0+Y_l\right)H_{f_{il}}(u)   }  \times \nonumber \\ 
& & g_0(Y_{0}) g_1(Y_{1}) \cdots g_J(Y_J)  dY_{0}dY_{1}\cdots dY_J du \nonumber \\
&=& \int_{0}^{t} h_{f_{ij}}(u)  \prod_{l \ne j}  \int_0^\infty  e^{  - H_{f_{il}}(u) Y_l}  g_l(Y_{l}) dY_l \times \nonumber \\
& & \bigg [  \frac{\omega_0}{\omega_j}  \int_{0}^{\infty} Y_0 e^{  - \sum{l=1}^J \left \{ \frac{\omega_0}{\omega_l} H_{f_{il}}(u)  \right \} Y_0} g_0(Y_{0}) dY_0  
  \int_{0}^{\infty} e^{  -  H_{f_{ij}}(u) Y_j} g_j(Y_{j}) dY_j  +  \nonumber \\
& &  \int_{0}^{\infty}  e^{  - \sum_{l=1}^J \left \{ \frac{\omega_0}{\omega_l} H_{f_{il}}(u)  \right \} Y_0} g_0(Y_{0}) dY_0   
\int_{0}^{\infty} Y_j e^{  -  H_{f_{ij}}(u) Y_j} g_j(Y_{j}) dY_j  \bigg ]  du \nonumber \\
&=& \int_{0}^{t} h_{f_{ij}}(u )  \prod_{l\ne j} \phi_l \left\{ H_{f_{il}}(u) \right \}  
 \bigg [  \frac{\omega_0}{\omega_j}   (-1)  \phi_0^{(1)} \left\{ \sum_{l=1}^J \frac{\omega_0}{\omega_l} H_{f_{il}}(u)  \right \} 
  \phi_j \{  H_{f_{ij}}(u) \}   +  \nonumber \\
& &   \phi_0 \left\{    \sum_{l=1}^J \frac{\omega_0}{\omega_l} H_{f_{il}}(u)  \right \} 
  (-1) \phi_j^{(1)} \{  H_{f_{ij}}(u) \}   \bigg ]  du \nonumber \\
&=& \int_{0}^{t} h_{f_{ij}}(u) \prod_{l\ne j} \bigg( 1+ \frac{H_{f_{il}}(u)}{\omega_l}\bigg)^{-k_l}  \bigg [ 
\frac{\omega_0}{\omega_j}   \left\{  1 + \sum_{l=1}^J \frac{H_{f_{il}}(u)}{\omega_l}  \right \}^{-k_0-1}  \left \{ 1 +  \frac{H_{f_{ij}}(u)}{\omega_j} \right \}^{-k_j} + \nonumber \\
& &  \left\{  1 +  \sum_{l=1}^J \frac{H_{f_{il}}(u)}{\omega_l}  \right \}^{-k_0} \frac{k_j}{\omega_j}  
        \left \{ 1 +  \frac{H_{f_{ij}}(u)}{\omega_j} \right \}^{-k_j -1}\bigg ]  du \nonumber \\
&= & \int_{0}^{t} h_{f_{i1}}(u) \prod_{l \ne j}  \left\{ 1+ \frac{H_{f_{il}}(u)}{\omega_l}\right\} ^{-k_l}  \left \{ 1 +  \frac{H_{f_{ij}}(u)}{\omega_j} \right \}^{-k_j}  
\left\{  1 + \sum_{l=1}^J \frac{H_{f_{il}}(u)}{\omega_l}  \right \}^{-k_0}  \times \nonumber \\
& & \bigg [ \frac{k_0}{\omega_j}  \left\{  1 + \sum_{l=1}^J  \frac{H_{f_{il}}(u)}{\omega_l}   \right \}^{-1} + 
 \frac{k_j}{\omega_j}  \left \{ 1 +  \frac{H_{f_{ij}}(u)}{\omega_j} \right \}^{-1}\bigg ]  du   \label{eqn:pen}                              
\end{eqnarray}
where the covariate history $\bm{X}_{f_{i}} (u) $ is removed from the hazard and cumulative hazard functions for simplicity and calculation details for PE, ED and CO models are specified in Web Appendix B.

\subsection{Variance Estimation}

The variance-covariance matrix of $\bm{\hat{\theta}}$ is estimated using a robust sandwich variance estimator, 
$$ V(\bm{\hat{\theta}})=I_o(\bm{\theta})^{-1} J(\bm{\theta}) I_o(\bm{\theta})^{-1}, $$
where $I_o(\bm{\theta})$ is the observed information matrix and $J(\bm{\theta})$ is the expected information matrix. They can be obtained by  
\begin{eqnarray*}
I_o(\bm{\theta})&=& -\frac{\partial^2 \ell_C(\bm{\theta}) }{\partial \bm{\theta}^{T} \partial \bm{\theta} }  \\
J(\bm{\theta})&=& \sum_f U_f(\bm{\theta})U_f^\top(\bm{\theta}) \\
U_f(\bm{\theta})&=&  \frac{\partial \mathrm{log}L_f(\bm{\theta}) }{\partial \bm{\theta}} -  \frac{\partial \mathrm{log}A_f(\bm{\theta}) }{\partial \bm{\theta}} . \nonumber
\end{eqnarray*}

The variance estimates $\hat{V}(\bm{\hat{\theta}})$ are obtained by evaluating $I_o(\bm{\theta})$ and $J(\bm{\theta})$ at the maximum-likelihood estimate $\bm{\hat{\theta}}$.

The robust variance estimator for the cause-specfic penetrance estimate, $F_j(t|\bm{\hat{\theta}})$, is obtained using Delta method:
$$ V(F_j(t|\bm{\hat{\theta}}))= D^\top_{\bm{\theta}}(t) V(\bm{\hat{\theta}}) D_{\bm{\theta}}(t), $$
where  $D_{\bm{\theta}}(t)$ is the vector of partial derivatives of $F_j(t|\bm{\theta})$ with respect to $\bm{\theta}$.  The variance estimates $\hat{V}(F_j(t|\bm{\hat{\theta}}))$ are obtained by plugging in $\bm{\hat\theta}$. 

\section{Simulation study}  

\subsection{Simulation Study Design}
We conducted simulation studies to assess the finite-sample properties of our proposed approach. We considered $J=2$ competing events with a TVC affecting a single event. Our simulated datasets mimic \textit{BRCA1} mutation positive families from the Breast Cancer Family Registry (BCFR) used in our application with respect to family structure and inclusion criteria. True parameter values were obtained after fitting our model to the real data. For each dataset, 500 families were generated under PE, ED and CO TVC models, each with low, medium and high familial dependence, which corresponds to $k_1=7$ ($\tau=0.07$), $3.5$ ($\tau=0.13$) and $1$ ($\tau=0.33$), respectively, where $\tau$ represents a Kendall's tau. A value close to 1 indicates higher dependence among the family relatives' failure times. The parameter $k_2$ was fixed at the estimated value obtained from the real data analysis. We consider the situation where $k_0$ goes to zero, i.e., independent frailties, as in our real data analysis, the parameters associated with the TVCs and penetrance functions (which are our main interests in these simulations) were not very sensitive to the presence of correlation between the frailties. All combinations of parameters can be found in Table \ref{t:one}. The model included a mutation status as a TIC affecting both events and a TVC, which can be either MS or RRSO, for event 1. Detailed steps of data generation are presented in Web Appendix C. For each scenario, the model parameters and penetrance estimators are evaluated based on 500 simulations by comparing bias, empirical standard error (ESE), average standard error (ASE) and empirical coverage probability (ECP). Bias is defined as the difference between mean estimate, $\bar{\hat{\beta}}$ and the true value of the parameter, $\beta$; ESE is obtained by the standard deviation of the estimates over all simulations, $\sqrt{ \sum_{i=1}^B(\hat{\beta_i}-\bar{\hat{\beta}})^2/(B-1)},$ where $B=500$ is the number of simulations and $\hat \beta_i$ is the parameter estimate from simulation $i$, $i=1, \ldots, B$ and $\bar{\hat{\beta}}$ is the average of the estimates from $B$ simulations; ASE is obtained by $\sum_{i=1}^B SE(\hat{\beta_i})/B$, the average of robust standard errors (SEs) from each simulation. Finally, ECP is the proportion of times 95\% confidence interval (CI) defined as $\hat{\beta_i} \pm Z_{0.975}$SE($\hat{\beta_i}$) include true value $\beta$ for $i=1,\ldots,B$.

In addition, we also investigated the robustness of the proposed model to the misspecification of TVC function in our simulations. Bias and efficiency of the misspecified TVC function are evaluated in comparison to the true TVC model.  Simulations results based on $n=500$ families are presented below while Tables S1 and S2 include simulation results for $n=1000$ families.

\subsection{Simulation Results}

The simulation results for the model parameter estimates are summarized in Table \ref{t:one}. 
Biases of the parameter estimates related to the baseline hazard function ($\rho_1, \lambda_1, \rho_2, \lambda_2$) and regression coefficients ($\beta_{1tvc}, \beta_{1gene}, \beta_{2gene}$) are negligible across all the TVC models and the levels of familial dependences. ASEs and ESEs are very close to each other and ECPs are within acceptable range, i.e., between 0.93 and 0.97. The frailty parameter estimates are more biased especially for event 2 and their ECP is lower than the nominal level, 0.95 (ranged between 0.80 and 0.90). We also observed that ASEs tend to be larger than ESEs in the CO model. Coverage probability for $k_1$ was better than for $k_2$ and the bias decreases with the level of familial dependence.

Table \ref{t:two} summarizes the simulation results related to the penetrance estimators. While frailty parameter estimators suffer from bias, penetrance estimators by age 70 for both event 1, $F_1(70; \bm{X})$, and event 2, $F_2(70; \bm{X})$, performed well. The bias was negligible ($<$ 1\%) and the ECPs were close to the 0.95 nominal level and within acceptable range (between 0.93 and 0.97) regardless of the level of familial dependence. ASEs and ESEs agree with each other in PE model but ASEs tend to be slightly higher than ESEs in the ED and CO models.

Additional simulations were conducted to evaluate the robustness of the proposed model to misspecification of the TVC function. We generated datasets under each TVC model assumption considering a medium familial dependence level ($k_1=3.5$) and then fitting the wrong TVC models to them. Tables S3 and S4 summarize  the simulation results for penetrance estimates under TVC misspecification. 
As expected, fitting ED and CO models on the dataset generated under a PE TVC leads to minimal biases. However, we note that the coefficient $\beta_{1tvc}$ of a TVC is largely biased under the CO model. Table S3 shows the TVC effect $\beta_{1tvc}$ is underestimated while $\eta_0$ is overestimated. The overall effect on penetrance is however unbiased since the bias on these two parameters is in opposite direction. Fitting a CO model on ED-generated data does not result in any bias. In other situations where  a simpler TVC model is fitted to more complex true TVC models, substantial biases are observed for the individuals with TVC = 1. Therefore, in practice, it is necessary to fit all three models and select the best model according to the lowest AIC values. In our simulations we note that the correct model is selected about 88\% of the time with this selection criteria. In Tables S1 and S2, we present additional simulation results for parameter and penetrance estimators for a larger number of families $n=1000$. In brief, when $n=1000$ the bias is substantially lower for all parameters, especially the frailty parameters, and their ECPs greatly improve (0.88 $\sim$ 0.93 for $k_2$). Similarly, penetrance estimators are less biased, i.e. less than 0.1\%.      

\begin{table}
\caption{Empirical parameter estimates from the competing risks model with a time varying covariate (TVC) under low ($k_1=7$), medium ($k_1=3.5$) and high ($k_1=1$) familial dependence;  permanent exposure (PE), exponential decay (ED) or Cox and Oaks (CO) models are considered for TVC. For each scenario, the mean bias, empirical standard error (ESE), average standard error (ASE) and estimated 95\% coverage probability (ECP) are obtained from 500 replicates each with $n=500$ families. }
\begin{adjustbox}{width=1\textwidth,center=\textwidth}
{\begin{tabular}{llrrrrrlrrrrrlrrrrr}
\hline
\multicolumn{1}{l}{TVC} & \multicolumn{1}{c}{} & \multicolumn{1}{c}{True} & \multicolumn{4}{c}{$k_1=7$, $\tau=0.07$} & \multicolumn{1}{c}{} & \multicolumn{1}{c}{True} & \multicolumn{4}{c}{$k_1=3.5$, $\tau=0.13$} & \multicolumn{1}{c}{} & \multicolumn{1}{c}{True} & \multicolumn{4}{c}{$k_1=1$, $\tau=0.33$} \\ \cline{4-7} \cline{10-13} \cline{16-19} 
\multicolumn{1}{l}{model} & \multicolumn{1}{c}{} & \multicolumn{1}{c}{value} & \multicolumn{1}{c}{Bias} & \multicolumn{1}{c}{ESE} & \multicolumn{1}{c}{ASE} & \multicolumn{1}{c}{ECP} & \multicolumn{1}{c}{} & \multicolumn{1}{c}{value} & \multicolumn{1}{c}{Bias} & \multicolumn{1}{c}{ESE} & \multicolumn{1}{c}{ASE} & \multicolumn{1}{c}{ECP} & \multicolumn{1}{c}{} & \multicolumn{1}{c}{value} & \multicolumn{1}{c}{Bias} & \multicolumn{1}{c}{ESE} & \multicolumn{1}{c}{ASE} & \multicolumn{1}{c}{ECP} \\ \hline
PE & $\mathrm{log}(\lambda_1)$ & -4.83 & -0.01 & 0.06 & 0.06 & 0.95 &  & -4.83 & 0.00 & 0.06 & 0.06 & 0.95 &  & -4.83 & 0.00 & 0.06 & 0.06 & 0.94 \\
 & $\mathrm{log}(\rho_1)$ & 0.88 & 0.00 & 0.03 & 0.03 & 0.94 &  & 0.88 & 0.00 & 0.03 & 0.03 & 0.93 &  & 0.88 & 0.00 & 0.03 & 0.03 & 0.96 \\
 & $\mathrm{log}(\lambda_2)$ & -4.96 & -0.01 & 0.09 & 0.10 & 0.95 &  & -4.96 & -0.02 & 0.10 & 0.10 & 0.94 &  & -4.96 & -0.01 & 0.09 & 0.10 & 0.96 \\
 & $\mathrm{log}(\rho_2)$ & 1.12 & 0.00 & 0.07 & 0.07 & 0.95 &  & 1.12 & 0.00 & 0.07 & 0.07 & 0.95 &  & 1.12 & 0.00 & 0.06 & 0.07 & 0.96 \\
 & $\beta_{1gene}$ & 1.95 & 0.01 & 0.12 & 0.12 & 0.95 &  & 1.95 & 0.01 & 0.12 & 0.12 & 0.96 &  & 1.95 & 0.00 & 0.12 & 0.11 & 0.94 \\
 & $\beta_{2gene}$ & 1.19 & 0.03 & 0.23 & 0.23 & 0.96 &  & 1.19 & 0.03 & 0.24 & 0.23 & 0.95 &  & 1.19 & 0.02 & 0.22 & 0.24 & 0.96 \\
 & $\beta_{1tvc}$ & 0.67 & 0.01 & 0.11 & 0.11 & 0.95 &  & 0.67 & 0.00 & 0.10 & 0.11 & 0.96 &  & 0.67 & 0.00 & 0.11 & 0.11 & 0.96 \\
 & $\mathrm{log}(k_1)$ & 1.95 & 0.24 & 1.08 & 0.85 & 0.92 &  & 1.25 & 0.13 & 0.69 & 0.48 & 0.95 &  & 0.00 & 0.02 & 0.25 & 0.25 & 0.95 \\
 & $\mathrm{log}(k_2)$ & 1.06 & 0.62 & 2.17 & 1.38 & 0.80 &  & 1.06 & 0.72 & 2.20 & 1.41 & 0.84 &  & 1.06 & 0.61 & 2.05 & 1.46 & 0.86 \\
 &  &  &  &  &  &  &  &  &  &  &  &  &  &  &  &  &  &  \\
ED & $\mathrm{log}(\lambda_1)$ & -4.83 & -0.01 & 0.05 & 0.06 & 0.96 &  & -4.83 & 0.00 & 0.06 & 0.06 & 0.95 &  & -4.83 & 0.00 & 0.06 & 0.06 & 0.96 \\
 & $\mathrm{log}(\rho_1)$ & 0.83 & 0.00 & 0.03 & 0.03 & 0.96 &  & 0.83 & 0.00 & 0.03 & 0.03 & 0.95 &  & 0.83 & 0.00 & 0.03 & 0.03 & 0.96 \\
 & $\mathrm{log}(\lambda_2)$ & -4.96 & 0.00 & 0.09 & 0.09 & 0.95 &  & -4.96 & -0.01 & 0.09 & 0.09 & 0.96 &  & -4.96 & -0.01 & 0.09 & 0.09 & 0.95 \\
 & $\mathrm{log}(\rho_2)$ & 1.08 & 0.00 & 0.06 & 0.06 & 0.95 &  & 1.08 & 0.00 & 0.06 & 0.06 & 0.95 &  & 1.08 & 0.00 & 0.06 & 0.06 & 0.95 \\
 & $\beta_{1gene}$ & 1.86 & 0.00 & 0.12 & 0.12 & 0.96 &  & 1.86 & 0.01 & 0.11 & 0.12 & 0.95 &  & 1.86 & 0.01 & 0.11 & 0.11 & 0.94 \\
 & $\beta_{2gene}$ & 1.22 & 0.01 & 0.20 & 0.21 & 0.95 &  & 1.22 & 0.03 & 0.22 & 0.21 & 0.96 &  & 1.22 & 0.02 & 0.21 & 0.22 & 0.96 \\
 & $\beta_{1tvc}$ & 1.87 & 0.03 & 0.25 & 0.25 & 0.94 &  & 1.87 & -0.01 & 0.25 & 0.25 & 0.95 &  & 1.87 & 0.03 & 0.24 & 0.24 & 0.94 \\
 & $\mathrm{log}(\eta)$ & -1.28 & 0.02 & 0.32 & 0.31 & 0.94 &  & -1.28 & 0.00 & 0.32 & 0.31 & 0.94 &  & -1.28 & 0.03 & 0.30 & 0.30 & 0.94 \\
 & $\mathrm{log}(k_1)$ & 1.95 & 0.23 & 0.99 & 0.88 & 0.93 &  & 1.25 & 0.08 & 0.49 & 0.48 & 0.97 &  & 0.00 & 0.02 & 0.23 & 0.24 & 0.96 \\
 & $\mathrm{log}(k_2)$ & 1.18 & 0.51 & 2.04 & 1.18 & 0.85 &  & 1.18 & 0.53 & 1.70 & 1.26 & 0.84 &  & 1.18 & 0.48 & 1.47 & 1.28 & 0.84 \\
 &  &  &  &  &  &  &  &  &  &  &  &  &  &  &  &  &  &  \\
CO & $\mathrm{log}(\lambda_1)$ & -4.83 & 0.00 & 0.05 & 0.06 & 0.95 &  & -4.83 & 0.00 & 0.05 & 0.06 & 0.94 &  & -4.83 & 0.00 & 0.05 & 0.06 & 0.96 \\
 & $\mathrm{log}(\rho_1)$ & 0.83 & 0.00 & 0.03 & 0.03 & 0.94 &  & 0.83 & 0.00 & 0.03 & 0.03 & 0.96 &  & 0.83 & 0.00 & 0.03 & 0.03 & 0.97 \\
 & $\mathrm{log}(\lambda_2)$ & -4.96 & 0.00 & 0.07 & 0.09 & 0.95 &  & -4.96 & 0.00 & 0.07 & 0.09 & 0.97 &  & -4.96 & 0.00 & 0.08 & 0.09 & 0.95 \\
 & $\mathrm{log}(\rho_2)$ & 1.07 & 0.00 & 0.05 & 0.06 & 0.96 &  & 1.07 & 0.00 & 0.05 & 0.06 & 0.97 &  & 1.07 & 0.00 & 0.05 & 0.06 & 0.96 \\
 & $\beta_{1gene}$ & 2.08 & 0.01 & 0.10 & 0.12 & 0.94 &  & 2.08 & 0.01 & 0.10 & 0.12 & 0.95 &  & 2.08 & 0.01 & 0.09 & 0.11 & 0.96 \\
 & $\beta_{2gene}$ & 1.57 & 0.00 & 0.17 & 0.21 & 0.98 &  & 1.57 & 0.00 & 0.17 & 0.21 & 0.94 &  & 1.57 & 0.01 & 0.16 & 0.21 & 0.97 \\
 & $\beta_{1tvc}$ & 1.52 & 0.04 & 0.32 & 0.42 & 0.96 &  & 1.52 & 0.04 & 0.33 & 0.42 & 0.94 &  & 1.52 & 0.02 & 0.32 & 0.42 & 0.96 \\
 & $\mathrm{log}(\eta)$ & -0.18 & -0.02 & 0.50 & 0.58 & 0.90 &  & -0.18 & 0.01 & 0.50 & 0.60 & 0.91 &  & -0.18 & -0.03 & 0.48 & 0.62 & 0.91 \\
 & $\eta_0$ & 0.21 & -0.02 & 0.12 & 0.14 & 0.95 &  & 0.21 & -0.01 & 0.12 & 0.14 & 0.96 &  & 0.21 & -0.02 & 0.12 & 0.14 & 0.95 \\ 
 & $\mathrm{log}(k_1)$ & 1.95 & 0.20 & 0.74 & 0.86 & 0.91 &  & 1.25 & 0.10 & 0.39 & 0.46 & 0.96 &  & 0.00 & 0.02 & 0.18 & 0.22 & 0.97 \\
 & $\mathrm{log}(k_2)$ & 1.26 & 0.38 & 1.15 & 1.39 & 0.86 &  & 1.26 & 0.35 & 0.98 & 1.40 & 0.90 &  & 1.26 & 0.36 & 1.10 & 1.32 & 0.87 \\ \hline
\multicolumn{19}{l}{$\lambda_j$ and $\rho_j$ are baseline hazard parameters for event $j, j=1,2$; }\\
\multicolumn{19}{l}{$\beta_{jgene}$ is the regression coefficient of a time-invariant covariate for event $j$;}\\
\multicolumn{19}{l}{$\beta_{1tvc}$, $\eta$ and $\eta_0$ are parameters to describe TVC effects; $k_j$ is the frailty parameter for event $j$.  }\\ 
\end{tabular}}
\end{adjustbox}
\bigskip
\label{t:one}
\end{table}

\begin{table}
\caption{Empirical penetrance estimates by age 70 for the competing risks model with a time varying covariate (TVC) under low ($k_1=7$), medium ($k_1=3.5$) and high ($k_1=1$) familial dependence;  permanent exposure (PE), exponential decay (ED) or Cox and Oaks (CO) models are considered for TVC; $F_1(70; \mbox{TVC, G})$ and $F_2(70; \mbox{TVC, G})$ are cause-specific penetrance estimators (\%) by age 70 for event 1 and event 2, respectively, given TVC and mutation status (G), and TVC occurred at age 35 if $\mbox{TVC}=1$. For each scenario, the mean bias, empirical standard error (ESE), average standard error (ASE) and estimated 95\% coverage probability (ECP) are obtained from 500 replicates each with $n=500$ families. 
}

\label{t:two}

\begin{adjustbox}{width=1\textwidth,center=\textwidth}
{\begin{tabular}{llrrrrrlrrrrrlrrrrr}
\hline
\multicolumn{1}{l}{TVC} & \multicolumn{1}{c}{} & \multicolumn{1}{c}{True} & \multicolumn{4}{c}{$k_1=7$, $\tau=0.07$} & \multicolumn{1}{c}{} & \multicolumn{1}{c}{True} & \multicolumn{4}{c}{$k_1=3.5$, $\tau=0.13$} & \multicolumn{1}{c}{} & \multicolumn{1}{c}{True} & \multicolumn{4}{c}{$k_1=1$, $\tau=0.33$} \\ \cline{4-7} \cline{10-13} \cline{16-19} 
\multicolumn{1}{l}{model} & \multicolumn{1}{c}{} & \multicolumn{1}{c}{value} & \multicolumn{1}{c}{Bias} & \multicolumn{1}{c}{ESE} & \multicolumn{1}{c}{ASE} & \multicolumn{1}{c}{ECP} & \multicolumn{1}{c}{} & \multicolumn{1}{c}{value} & \multicolumn{1}{c}{Bias} & \multicolumn{1}{c}{ESE} & \multicolumn{1}{c}{ASE} & \multicolumn{1}{c}{ECP} & \multicolumn{1}{c}{} & \multicolumn{1}{c}{value} & \multicolumn{1}{c}{Bias} & \multicolumn{1}{c}{ESE} & \multicolumn{1}{c}{ASE} & \multicolumn{1}{c}{ECP} \\ \hline
PE & $F_1(70; \mbox{TVC = 0, G = 0})$ & 12.56 & -0.10 & 1.38 & 1.36 & 0.95 &  & 12.45 & 0.01 & 1.33 & 1.40 & 0.94 &  & 11.93 & 0.07 & 1.48 & 1.45 & 0.94 \\
 & $F_1(70; \mbox{TVC = 1, G = 0})$ & 21.92 & -0.01 & 2.45 & 2.45 & 0.94 &  & 21.58 & 0.02 & 2.37 & 2.48 & 0.95 &  & 20.09 & 0.13 & 2.49 & 2.50 & 0.96 \\
 & $F_1(70; \mbox{TVC = 0, G = 1})$ & 56.52 & -0.33 & 3.20 & 3.18 & 0.94 &  & 54.51 & 0.12 & 3.39 & 3.42 & 0.94 &  & 46.80 & -0.02 & 3.84 & 3.92 & 0.95 \\
 & $F_1(70; \mbox{TVC = 1, G = 1})$ & 75.63 & -0.23 & 3.75 & 3.74 & 0.94 &  & 72.59 & 0.03 & 4.08 & 4.06 & 0.94 &  & 61.08 & -0.04 & 4.61 & 4.79 & 0.94 \\
 &  &  &  &  &  &  &  &  &  &  &  &  &  &  &  &  &  &  \\
 & $F_2(70; \mbox{TVC = 0, G = 0})$ & 4.73 & -0.08 & 0.82 & 0.85 & 0.94 &  & 4.73 & -0.08 & 0.87 & 0.85 & 0.93 &  & 4.74 & -0.05 & 0.79 & 0.88 & 0.95 \\
 & $F_2(70; \mbox{TVC = 1, G = 0})$ & 4.45 & -0.08 & 0.77 & 0.80 & 0.94 &  & 4.45 & -0.08 & 0.82 & 0.80 & 0.93 &  & 4.49 & -0.05 & 0.75 & 0.83 & 0.95 \\
 & $F_2(70; \mbox{TVC = 0, G = 1})$ & 9.68 & 0.04 & 1.16 & 1.15 & 0.94 &  & 9.85 & -0.04 & 1.16 & 1.18 & 0.95 &  & 10.52 & 0.02 & 1.29 & 1.28 & 0.95 \\
 & $F_2(70; \mbox{TVC = 1, G = 1})$ & 7.12 & 0.01 & 0.91 & 0.89 & 0.94 &  & 7.42 & -0.04 & 0.91 & 0.92 & 0.95 &  & 8.56 & 0.00 & 1.04 & 1.04 & 0.95 \\
 &  &  &  &  &  &  &  &  &  &  &  &  &  &  &  &  &  &  \\
ED & $F_1(70; \mbox{TVC = 0, G = 0})$ & 13.55 & -0.05 & 1.39 & 1.42 & 0.94 &  & 13.42 & -0.02 & 1.41 & 1.44 & 0.94 &  & 12.82 & -0.04 & 1.47 & 1.47 & 0.94 \\
 & $F_1(70; \mbox{TVC = 1, G = 0})$ & 15.49 & 0.03 & 1.64 & 1.64 & 0.94 &  & 15.32 & 0.05 & 1.62 & 1.66 & 0.94 &  & 14.54 & 0.00 & 1.61 & 1.68 & 0.97 \\
 & $F_1(70; \mbox{TVC = 0, G = 1})$ & 55.65 & -0.28 & 2.70 & 3.07 & 0.97 &  & 53.68 & -0.05 & 3.03 & 3.27 & 0.96 &  & 46.14 & 0.12 & 3.56 & 3.68 & 0.96 \\
 & $F_1(70; \mbox{TVC = 1, G = 1})$ & 60.49 & -0.10 & 2.99 & 3.33 & 0.97 &  & 58.24 & 0.10 & 3.26 & 3.54 & 0.97 &  & 49.69 & 0.21 & 3.67 & 3.94 & 0.96 \\
 &  &  &  &  &  &  &  &  &  &  &  &  &  &  &  &  &  &  \\
 & $F_2(70; \mbox{TVC = 0, G = 0})$ & 5.39 & 0.01 & 0.90 & 0.91 & 0.95 &  & 5.39 & -0.07 & 0.86 & 0.92 & 0.95 &  & 5.41 & -0.05 & 0.85 & 0.93 & 0.96 \\
 & $F_2(70; \mbox{TVC = 1, G = 0})$ & 5.26 & 0.01 & 0.87 & 0.89 & 0.95 &  & 5.26 & -0.07 & 0.83 & 0.89 & 0.95 &  & 5.28 & -0.06 & 0.83 & 0.91 & 0.95 \\
 & $F_2(70; \mbox{TVC = 0, G = 1})$ & 11.38 & 0.05 & 1.18 & 1.22 & 0.96 &  & 11.57 & 0.04 & 1.29 & 1.24 & 0.95 &  & 12.34 & -0.05 & 1.34 & 1.35 & 0.95 \\
 & $F_2(70; \mbox{TVC = 1, G = 1})$ & 9.97 & 0.01 & 1.05 & 1.09 & 0.96 &  & 10.22 & 0.01 & 1.12 & 1.12 & 0.95 &  & 11.20 & -0.07 & 1.20 & 1.22 & 0.95 \\
 &  &  &  &  &  &  &  &  &  &  &  &  &  &  &  &  &  &  \\
CO & $F_1(70; \mbox{TVC = 0, G = 0})$ & 13.54 & 0.02 & 1.36 & 1.42 & 0.95 &  & 13.41 & 0.02 & 1.44 & 1.43 & 0.95 &  & 12.81 & 0.08 & 1.34 & 1.43 & 0.96 \\
 & $F_1(70; \mbox{TVC = 1, G = 0})$ & 16.60 & -0.04 & 2.00 & 2.03 & 0.95 &  & 16.41 & -0.04 & 2.02 & 2.03 & 0.94 &  & 15.52 & -0.02 & 1.89 & 1.99 & 0.95 \\
 & $F_1(70; \mbox{TVC = 0, G = 1})$ & 61.12 & 0.07 & 2.90 & 2.93 & 0.96 &  & 58.82 & 0.25 & 3.15 & 3.10 & 0.94 &  & 50.11 & 0.32 & 3.32 & 3.49 & 0.97 \\
 & $F_1(70; \mbox{TVC = 1, G = 1})$ & 67.55 & -0.15 & 3.94 & 3.73 & 0.93 &  & 64.90 & 0.06 & 3.80 & 3.86 & 0.95 &  & 54.88 & 0.09 & 3.94 & 4.09 & 0.95 \\
 &  &  &  &  &  &  &  &  &  &  &  &  &  &  &  &  &  &  \\
 & $F_2(70; \mbox{TVC = 0, G = 0})$ & 5.53 & 0.04 & 0.87 & 0.93 & 0.95 &  & 5.53 & 0.05 & 0.88 & 0.93 & 0.95 &  & 5.55 & -0.02 & 0.89 & 0.95 & 0.95 \\
 & $F_2(70; \mbox{TVC = 1, G = 0})$ & 5.39 & 0.03 & 0.85 & 0.90 & 0.95 &  & 5.39 & 0.04 & 0.85 & 0.91 & 0.95 &  & 5.42 & -0.02 & 0.87 & 0.93 & 0.95 \\
 & $F_2(70; \mbox{TVC = 0, G = 1})$ & 14.27 & -0.06 & 1.24 & 1.37 & 0.98 &  & 14.61 & -0.02 & 1.38 & 1.41 & 0.94 &  & 15.91 & -0.08 & 1.51 & 1.55 & 0.95 \\
 & $F_2(70; \mbox{TVC = 1, G = 1})$ & 12.35 & -0.06 & 1.22 & 1.28 & 0.96 &  & 12.77 & -0.02 & 1.31 & 1.32 & 0.95 &  & 14.36 & -0.07 & 1.41 & 1.45 & 0.95 \\ \hline
\end{tabular}}
\end{adjustbox}
\bigskip
\label{t:two}
\end{table}

\section{Application to \textit{BRCA1} Families from BCFR}
\label{s:inf}

\subsection{Data}

Our analyses focus on \textit{BRCA1} carrier families recruited through the BCFR \cite{John04}. The BCFR was established in 1995 with six participating sites from the USA, Australia and Canada including Ontario Cancer Care. It enrolled most of the families from 1996 to 2000 while continuing to recruit additional families satisfying its criteria, i.e., families were included whenever they segregate \textit{BRCA1} or \textit{BRCA2} mutations, exhibit multiple cases of breast or ovarian cancer, are Ashkenazi Jewish ancestry or from specific racial and ethnic groups. For the population-based families, each family includes the proband, i.e. the initial member of the family to be identified, as well as the first and the second degree relatives. The data have extensive information on the family members including the ages of the breast/ovarian cancer diagnosis, study entry, RRSO uptake, mammographic screening and mutation status in \textit{BRCA1/2} gene. We restricted our data analyses to the \textit{BRCA1} families in the BCFR, which were identified from 498 probands including a total of 2,650 individuals. A complete description of the families is given in Table~\ref{t:descriptive}. 

\begin{table}[h!]
\caption{Characteristics of 498 \textit{BRCA1} positive families from the BCFR} \label{t:descriptive}
\begin{adjustbox}{width=1\textwidth}
\centering
\begin{tabular}{l r r r r r}
\hline
& Breast Cancer & Ovarian Cancer & Death & Unaffected & Total \\ \hline
{\textit{\textbf N(\%)}} & 924 (34.9\%) & 182 (6.9\%) & 958 (36.2\%) & 586 (22.1\%) & 2650 \\
{\textit{\textbf N(\%) of probands}} & 391 (78.5\%) & 43 (8.6\%) & 5 (1.0\%) & 59 (11.9\%) & 498 \\
{\textit{\textbf N(\%) of probands}} & \\ 
$\quad\quad$ {\textit{ at study entry}} & 386 (77.5\%) & 31 (6.2\%) & 0 (0\%) & 81 (16.3\%) & 498 \\
 &  &  &  &  &  \\
\textit{\textbf{Event age}} &  &  &  &  &  \\
mean (SD) & 44.2 (12.0) & 53.0 (11.5) & 70.5 (17.9) & 50.9 (16.2) & 55.8 (19.1) \\
min, max & 21.0, 86.0 & 28.0, 89.0 & 18.5, 102.5 & 18.1, 95.0 & 18.1, 102.5 \\
 &  &  &  &  &  \\
\textit{\textbf{BRCA1 mutation status}} &  &  &  &  &  \\
Noncarrier & 29 (3.1\%) & 4 (2.2\%) & 14 (1.5\%) & 229 (39.1\%) & 276 (10.4\%) \\
Carrier & 483 (52.3\%) & 55 (30.2\%) & 16 (1.7\%) & 192 (32.8\%) & 746 (28.2\%) \\
Untested & 412 (44.6\%) & 123 (67.6\%) & 928 (96.9\%) & 165 (28.2\%) & 1628 (61.4\%) \\
 &  &  &  &  &  \\
\multicolumn{2}{l}{\textit{\textbf{\# of mammographic screening}} } &  &  &  &  \\
0 & 722 (78.1\%) & 158 (86.8\%) & 944 (98.5\%) & 257 (43.9\%) & 2081 (78.5\%) \\
1 & 160 (17.3\%) & 19 (10.4\%) & 7 (0.7\%) & 174 (29.7\%) & 360 (13.6\%) \\
2 & 31 (3.4\%) & 4 (2.2\%) & 3 (0.3\%) & 63 (10.8\%) & 101 (3.8\%) \\
3+ & 11 (1.2\%) & 1 (0.5\%) & 4 (0.4\%) & 92 (15.7\%) & 108 (4.1\%) \\
 &  &  &  &  &  \\
\textit{\textbf{RRSO}} & 28 (3.0\%) & 0 (0\%) & 9 (0.9\%) & 129 (22.0\%) & 166 (6.3\%) \\ \hline
\multicolumn{5}{l}{RRSO stands for risk-reducing bilateral salpingo-oophorectomy.}\\
\multicolumn{5}{l}{SD stands for standard deviation.}
\end{tabular}
\end{adjustbox}
\end{table}

\subsection{Analyses}
Our main event of interest is the time to a first primary BC while a first primary OC and death (from other causes than BC or OC) are considered as competing events in our analyses. The Weibull distribution was used to fit the cause-specific baseline hazard functions. Age is considered as the time scale, i.e. age at diagnosis for women with either BC or OC, and age at last follow-up or death for women free of BC and OC. 
Age at RRSO is our main TVC of interest while the successive MS events are assumed to be confounding TVCs.  We considered up to three possible MS events. Prophylactic bilateral mastectomy was considered as a censoring variable for BC. We only accounted for screening and surgery histories before any events of interest (BC, OC, death or censored). When the age at RRSO was less than one year from the age at BC onset, we considered that both events occurred at the same time and thus RRSO did not affect BC (n = 12). The proportion of individuals with OC as first cancer is much lower than that of BC (6.9\% vs. 34.9\%). The proportion of women who underwent RRSO among the BC cohort is 3\%.

\subsection{Selection of the best TVC model}
For both RRSO and MS variables, we used the Akaike information criterion (AIC) to select the best TVC model and evaluated the three models, i.e. PE, ED and CO, for each of them. The best model corresponds to the CO model for both RRSO and the three MS-related variables with an AIC of 19077.43 (Table~\ref{tab:HR}).  The form of the hazard function corresponding the best model and that of other TVC models are displayed in Figure S1. The choice of the CO model means that for women with $BRCA1$ mutations, the effect of RRSO on BC reduces over time until reaching a threshold.

\subsection{Correlation between the competing events}
We found a significant correlation between the 2 competing events BC and OC conditional on the mutation status, estimated at 0.52 (95\% CI = 0.17 , 0.79) (see Method section). The variance of each frailty is 0.29 (se = 0.04) for BC and 0.40 (se = 0.13) for OC, corresponding to a Kendall's tau of 0.13 ( 95\% CI =  0.09,  0.20) and 0.17 (95\% CI =  0.11,  0.37), respectively, representing within familial correlation for each event. The correlation between BC and death and between OC and death was close to 0 and the frailty parameter corresponding time to death was not significant at the 5\% level. Therefore, we only considered the correlation between BC and OC in our final model, which involves the frailty parameters $k_0$, $k_1$ and $k_2$ in Table~\ref{t:model}.

\subsection{Effects of mutation status on the competing events, RRSO and MS on breast cancer}
The parameter estimates for the correlated competing risk models are given in Table~\ref{t:model}. The parameters $\beta_{1gene}$, $\beta_{2gene}$ and $\beta_{3gene}$ correspond to the $BRCA1$ mutation effect on the time to BC, OC and death, respectively. The 3 parameters are all significant at the 5\% level and yield hazard ratios of 9.53 (95\% CI = 7.44, 12.19), 4.41 (95\% CI = 2.81, 6.92) and 0.70 (95\% CI = 0.47, 0.81), respectively.  The other parameters $\beta_{1}$'s, $\eta$'s and $\eta_0$'s correspond to the 3 MS and RRSO effects at baseline, rates exponential decay and threshold values (see Method section). The RRSO and the 3 MSs were highly significant ($p < 0.001$) based on the likelihood ratio test when comparing a model with RRSO vs. no RRSO (the 3 MSs included) and a model with the 3 MSs vs. no MS (RRSO included), respectively. 

\begin{table}
\caption{Parameter estimates associated with BC in the {\it BRCA1} families from BCFR based on the model with competing risks (OC and death) assuming CO model for mammography screening and CO for risk reducing bilateral salpingo-oophorectomy.} \label{t:model}
\centering
\begin{tabular}{lrrr}\\ \hline 
 & \multicolumn{1}{c}{Estimate} & \multicolumn{1}{c}{SE} & \multicolumn{1}{c}{$p$-value} \\ \hline 

$\beta_{1gene}$ & 2.25 & 0.13 & $<0.01$\\
$\beta_{1_{MS1}}$ & 3.44 & 0.26 & $<0.01$ \\
$\beta_{1_{MS2}}$ & 3.97 & 0.46 & $<0.01$ \\
$\beta_{1_{MS3}}$ & 3.95 & 0.97 & $<0.01$ \\
  $\beta_{1_{RRSO}}$ & -1.79 & 0.71 & 0.01 \\
  $\log(\eta_{MS1})$ & 1.544 & 0.24 & $<0.01$ \\
  $\log(\eta_{MS2})$ & 0.87 & 0.37 & 0.02 \\
  $\log(\eta_{MS3})$ & 1.55 & 1.24 & 0.21 \\
  $\eta_{0_{MS1}}$ & 0.36 & 0.14 & 0.01 \\
  $\eta_{0_{MS2}}$ & -0.43 & 0.41 & 0.29 \\
  $\eta_{0_{MS3}}$ & -0.38 & 0.60 & 0.53 \\
  $\log(\eta_{RRSO})$ & -0.19 & 0.45 & 0.68 \\
  $\eta_{0_{RRSO}}$ & -0.41 & 0.24 & 0.08 \\
  $\log(k_1)$ & 0.63 & 0.41 & 0.12 \\
  $\log(k_2)$  & -0.04 & 0.79 & 0.96\\
  $\log(k_0)$  & 0.43 & 0.40 & 0.29\\
  $\beta_{2gene}$  & 1.48 & 0.23 & $<0.01$\\
  $\beta_{3gene}$  & -0.36 & 0.14 & 0.01\\
  \hline
  -loglik & 9514.72\\
  -loglik0$\dag$ & 9523.24\\
   p-value$\ast$ & $<$0.001\\ \hline 
   \multicolumn{4}{l}{ $\dag$ based on the  null model without RRSO}  \\
   \multicolumn{4}{l}{ $\ast$ testing for RRSO effect comparing to  }\\
   \multicolumn{4}{l}{ the null model using the likelihood ratio}\\
   \multicolumn{4}{l}{ test with df = 3}  \\
  & \\
\end{tabular} \\
\end{table}

\subsection{Time-dependent effect of RRSO on relative risk of BC in women with $BRCA1$ mutations}

The time-dependent association of the RRSO on BC can be assessed by its effect on the hazard function assessed by the hazard ratio (HR) given by $\exp\{\mu(x_{fi}(t))\}$ or on BC cumulative incidence (i.e., penetrance function), which are both defined as cause-specific functions. The time-dependent effect of RRSO was estimated on a continuous scale from 1 to 10 years after surgery (Table ~\ref{tab:HR}). Under the best fitting TVC model (i.e., the CO model) and assuming competing risks and MS adjustment, the overall effect of RRSO on BC risk is statistically significant in women with \textit{BRCA1} (p $<$ 0.001). Under this TVC model, the effect of RRSO reduces over time, i.e., HR = 0.30 (95\% CI = 0.09, 0.59) to HR = 0.66 (95\% CI = 0.42, 1.02)  from 1 to 10 years post surgery in \textit{BRCA1} mutation carriers. 

\begin{table}
\caption{Hazard ratios (and  their 95\% confidence intervals) measuring the time-dependent effect of risk-reducing salpingo-oophorectomy (RRSO) on BC risks
based on different TVC models (CO, Cox and Oakes; ED, exponential decay; PE, permanent exposure) in {\it BRCA1} families from the BCFR; 
Best TVC model for {\it BRCA1} families is indicated in bold. }\label{tab:HR}
\centering
\begin{tabular}{l c c c } 
\hline
Time & {\bf CO} & ED & PE \\ \hline 
1 & {\bf 0.30} &  0.28 & 0.55 \\
&{\bf (0.09, 0.59)} & (0.12, 0.69) &  (0.36, 0.82) \\
2 & {\bf 0.47} & 0.46 & 0.55 \\
& {\bf  (0.17, 0.77)} &  (0.20, 0.92) &  (0.36, 0.82) \\
3 & {\bf 0.57} &   0.63 & 0.55 \\
& {\bf (0.26, 0.85)} &  (0.28, 0.98) & (0.36, 0.82) \\
4 & {\bf  0.62} &  0.76 & 0.55 \\
 & {\bf (0.34, 0.91)} & (0.34, 1.00) & (0.36, 0.82) \\
 5 & {\bf 0.64} &  0.85 & 0.55 \\
 & {\bf (0.38, 0.94)} & (0.42, 1.00) & (0.36, 0.82) \\
6 & {\bf 0.66} &  0.90 & 0.55 \\
 & {\bf (0.40, 0.98)} & (0.48, 1.00) & (0.36, 0.82) \\
7 & {\bf 0.66} &  0.94 & 0.55 \\
& {\bf (0.41, 1.00)} & (0.55, 1.00) & (0.36, 0.82) \\
 8 & {\bf 0.66} &  0.96 & 0.55\\
& {\bf (0.41, 1.01)} &  (0.61, 1.00) & (0.36, 0.82) \\
 9 & {\bf 0.66} & 0.98 & 0.55 \\
& {\bf (0.41, 1.02)} &  (0.66, 1.00) & (0.36, 0.82) \\
10 & {\bf 0.66} & 0.99 & 0.55 \\
& {\bf (0.42, 1.02)} &  (0.71, 1.00) & (0.36, 0.82) \\ \hline 
LRT$^\dag$ &{\bf 17.04} & 12.89 & 10.09 \\
p-value & {\bf (0.001)} &  (0.005) & (0.018) \\
AIC &{\bf  19077.43} & 19079.58 & 19080.39 \\ \hline 
\multicolumn{4}{l} {$^\dag$ LRT, Likelihood ratio test statistics comparing to the null}\\
\multicolumn{4}{l} {model with no RRSO effect; }\\
\multicolumn{4}{l} {All models are adjusted for 3 MSs.}\\
\end{tabular}
\end{table}

\subsection{Time-dependent effect of RRSO on cumulative risk of BC among women with \textit{BRCA1} mutations}
\begin{figure}[!h]
\centering
\includegraphics[width=0.7\textwidth, height=0.5\textheight]{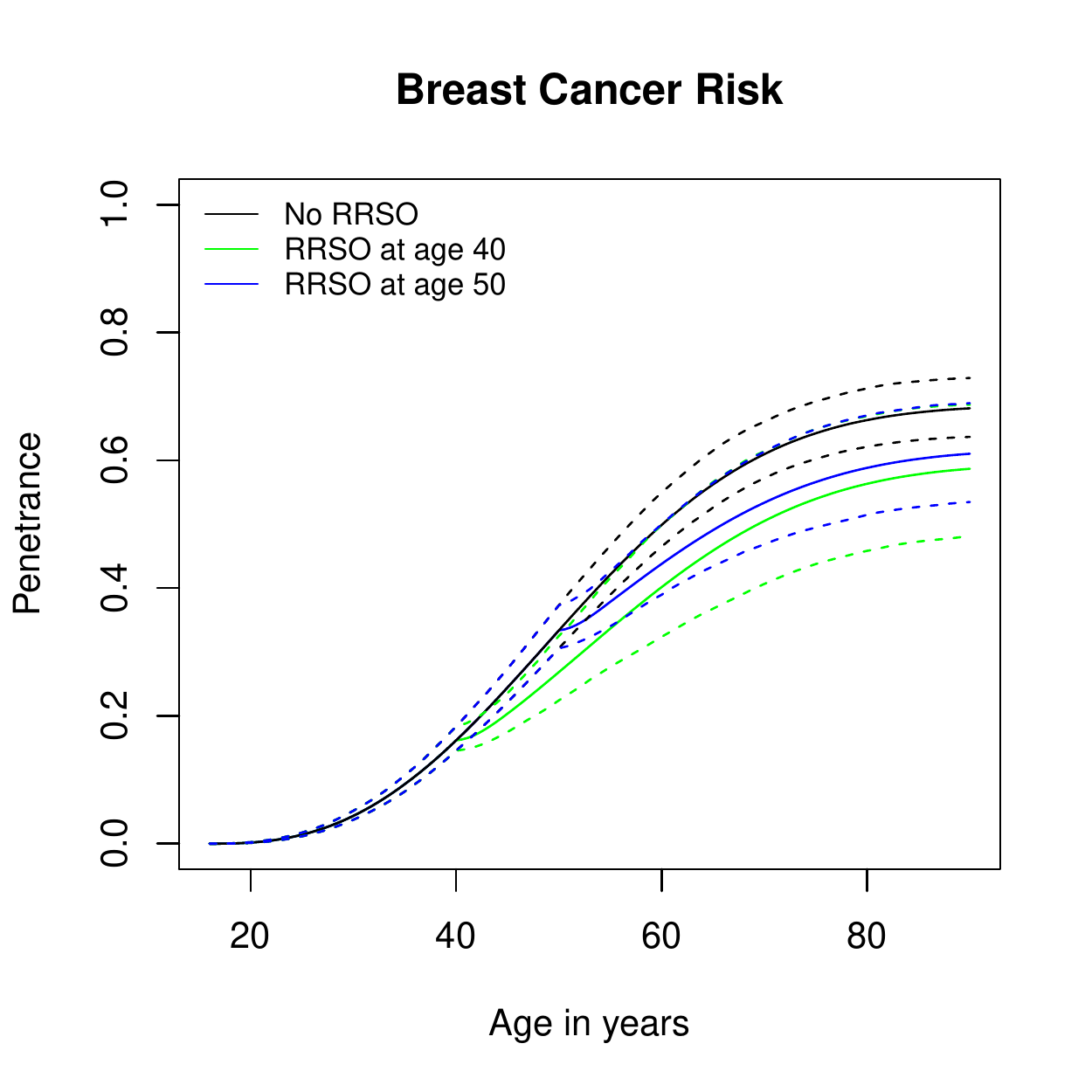} 
\caption{Breast cancer-specific penetrance estimates for mutation carriers with respect to risk-reducing salpingo-oophorectomy (RRSO) from the correlated competing-risks model. The black line represents a woman who did not have RRSO, the green line a woman who had RRSO at age 40 years and the blue line a woman who had RRSO at age 50 years. The dashed lines represent the 95\% confidence intervals.}\label{fig:penTVC}
\end{figure}

The cause-specific penetrance for BC for women without a RRSO is 61.0\% (95\% CI = 57.2, 66.0) by age 70 for women with a \textit{BRCA1} mutation and 12.0\% (95\% CI = 9.9, 14.2) for women within \textit{BRCA1} families but who do not carry a mutation (Figure~\ref{fig:penTVC} and Table~\ref{tab:penTVC}. The cause-specific penetrance of BC for a woman with RRSO at 40 years is 50.5\% (95\% CI = 40.6, 61.4) by age 70 for women with $BCRA1$ mutations (Figure~\ref{fig:penTVC} and Table~\ref{tab:penTVC}. For a woman with RRSO at 50 years, this penetrance is 53.4\% (95\% CI = 46.9, 61.3) while for a woman with RRSO at 30 years it is 49.0\% (95\% CI = 36.7, 62.3). 

\begin{table}
\caption{Penetrance estimates and their 95\% confidence intervals 
based on the best TVC model in the {\it BRCA1} from the BCFR}\label{tab:penTVC}
\centering
\begin{tabular}{l c c   } 
\hline
 & Age 50 & Age 70\\  \hline
 {\bf Breast cancer$^{\dag}$} & \\ 
$\quad$ Carriers & 33.4 \% (30.6, 37.3) & 61.0 \% (57.2, 66.0)\\
$\quad$ Non-carriers & 4.5\% (3.6, 5.5) & 12.0\% (9.9, 14.2)\\
$\quad$ RRSO at 30 years & 24.4\% (17.5, 33.5) & 49.0\% (36.7, 62.3)\\
$\quad$ RRSO at 35 years & 25.3\% (19.6, 32.5) & 49.6\% (38.3, 61.6)\\
$\quad$ RRSO at 40 years & 26.8\% (22.4, 32.5) & 50.5\% (40.6, 61.4)\\
$\quad$ RRSO at 50 years & 33.4\% (30.6, 37.3) & 53.4\% (46.9, 61.3)\\
 {\bf Ovarian cancer$^{\ddag}$} &   \\  
$\quad$ Carriers & 4.7\% (3.9, 6) & 11.2\% (9.1, 14.2)\\
$\quad$ Non-carriers & 1.4\% (1, 1.9) & 5.0\% (3.9, 6.6)\\
 \hline
 \multicolumn{3}{l}{$^{\dag}$Corresponds to a first breast cancer}\\
  \multicolumn{3}{l}{$^{\ddag}$Corresponds to a first ovarian cancer}\\
\end{tabular}
 \end{table}

\subsection{Sensitivity to RRSO modeling assumptions}

Our best TVC models assume a parametric form (exponential decay) for the variation of RRSO effect over time. To assess this assumption, we fitted a more general piece-wise TVC for RRSO, where the hazard ratio was constant within intervals but did not follow any particular functional form. We considered  three time intervals: $\le$ 2 years, 2-5 years and $>$ 5 years. The HR estimates from this model are close to the best TVC model and confirm that the exponential decay for RRSO effect over time is a reasonable assumption (data not shown). 

\subsection{Goodness-of-fit of the TVC model}
We evaluated the goodness-of-fit of our best TVC model using martingale residuals for each competing event, which are defined as the difference between the number of events of subject $i$ in family $f$ at time $T_{f_i}$ and the expected number of events computed by the cumulative hazard by the last observed time $T_{f_i}$. The martingale residuals are derived at both the individual level and the family level (Web Appendix D) and their martingale residuals plots are given in Figures S2--S4.
At both levels, their means are close to zero, indicating the good fit of the TVC model to the data.

\section{Discussion}
\label{s:discuss}

Members of {\it BRCA1} mutation positive families are exposed to a very high risk of developing BC or OC as first cancer and the risk of BC is likely to depend on time-varying covariates such as MS and RRSO in a complex manner. Most risk prediction models developed for these families do not account for competing risks nor for time-varying effects on BC. In this paper, we developed a flexible approach based on competing risks model, where the risk of the first competing event (BC) could depend on time-varying covariates. Our model provides cause-specific hazard functions and cumulative incidence functions that estimates age-specific risks of BC and OC, accounting for death as a competing event and residual familial correlation not due to the $BRCA1$ mutation segregating within the family. 

Our simulation studies demonstrate the good performances of our approach in terms of bias and precision of the estimators of model parameters  and cause-specific penetrances over different levels of familial correlations. The frailty-related parameter estimators had larger biases than other parameter estimators but these biases did not results in any biases of the cause-specific hazard functions and penetrances. This is a very important result since the cause-specific penetrance is used by genetic counsellors to guide clinical decisions such as prophylactic surgery or intensive screening for known mutation carriers or the decision to have genetic testing for unknown mutation carriers in {\it BRCA} families. Another important result is that, applying models with the wrong TVC function could also result in substantial biases of the parameter estimators when fitting a simpler model to a more complex time-varying function. It is therefore critical to select the correct TVC function to obtain accurate hazard ratio and cause-specific penetrance estimates. 

Our application to 498 {\it BRCA1} mutation positive families from the BCFR illustrates the importance of accounting for both competing risks and TVCs when estimating cause-specific penetrance of BC among mutation carriers. In addition, our results demonstrate the importance of the functional form of the TVC when assessing the role of RRSO on breast cancer, in line with our simulation results. In particular, under the best fitting TVC model (i.e., the CO model) with competing risks and MS adjustment, the overall effect of RRSO on BC risk was statistically significant in women with \textit{BRCA1} mutations. Under this TVC model, the effect of RRSO reduces over time, i.e., HR = 0.30 (95\% CI = 0.12, 0.69) to HR = 0.66 (95\% CI = 0.42, 1.02)  from 1 to 10 years post surgery in \textit{BRCA1} mutation carriers. In terms of cumulative risks, for a woman with RRSO at age 40 years, the cause-specific cumulative risk of BC was 50.5\% (95\% CI = 40.6, 61.4) by age 70 years for women with $BRCA1$ mutations compared with 61.0\% (95\% CI = 57.2, 66.0) for women without a RRSO. This result could have some importance for the clinical management of women carrying $BRCA1$ mutations but warrants further confirmation.

Our model assumes the TVCs as exogenous variables, i.e, the future path of the covariate is independent of the occurrence of BC \cite{Kalbfleisch02}, so that the hazard function at a specific time $t$ is influenced by the observed covariate history up to time $t$ in the regression model,. This assumption is realistic for prophylactic RRSO and scheduled MS in our application since the observation of RRSO and MS does not carry information about the status of BC; however, if the MS were performed in symptomatic women, the MS would not be exogenous since it could carry information about the status of BC. Even in that latter situation, our inference is based on the likelihood conditional on the covariate process up to the time $t$, so the future path of the covariate would not influence the occurrence of BC.

In the situation where the full path of the TVC is of research interest, e.g. even after the event of interest, some statistical approaches, such as the joint modeling of the TVCs as recurrent events and the cancer outcome as a terminal event, could be proposed. We have recently developed such approach for family data however it will require further extensions to be applicable to competing risks events \cite{Choi19}. 

Our model could also help evaluating more intervention options on BC risk, such as combinations of RRSO and MSs as well as the ages they could be introduced. It could be further extended to account for additional competing risks events, e.g. prophylactic mastectomy, and also to estimate the risks of successive cancer events after a first BC or OC, for example following our previous work \cite{Choi17}. Finally, we are planning to incorporate information on polygenic risk score from known genetic variants \cite{Kuchenbaecker17b}, that could modify BC and OC risks by incorporating a kinship matrix into  the cause-specific model for BC and/or OC \cite{Lakhal-Chaieb20}. These future developments should lead to a more comprehensive risk prediction model applicable to {\it BRCA} families as well as other families with increased genetic risks.

\section{Acknowledgements}
This work was supported by grant UM1 CA164920 from the USA National Cancer Institute. The content of this manuscript does not necessarily reflect the views or policies of the National Cancer Institute or any of the collaborating centers in the Breast Cancer Family Registry (BCFR), nor does mention of trade names, commercial products, or organizations imply endorsement by the USA Government or the BCFR. This research was also supported by two grants from the Canadian Institutes of Health Research (MOP 126186 \& 110053), an Interdisciplinary Health Research Team award from the Canadian Institutes of Health Research (Grant \# 43821), a grant from the Canadian Breast Cancer Foundation (BC-RG-15-2 competition), and Discovery Grants (\#RGPIN-2019-06549) from the Natural Sciences and Engineering Research Council of Canada.

\bibliographystyle{biorefs}
\bibliography{tvc_reference}

\clearpage

\newpage

\begin{center}
{\Huge \bf Online Supplementary Materials}
\end{center}


\section*{Web Appendix A:  Derivation of cumulative hazard function with a time-varying covariate}
For the time varying covariate $x_{f_i}(t)$ = 0 at $t<t_x$ and 1 at $t \geq t_x$, where $t_x$ is the time that change in value of time varying covariate occurred. 
The $j$th cause-specific cumulative hazard function with TVC for three TVC models (PE, ED, and CO) can be specified as
\begin{align*}
H_{f_{i}j}(t|X_{f_i}(t),z_{f_j}) \! &=  \! \int_0^{t} h_{0j}(u)z_{f_j}\mathrm{exp}\big\{ \mu(X_{f_i}(u))  \big\} du  \\
&=
\begin{cases}
    H_{0j}(t)z_{f_j}                                                        & \text{if } t < t_x \text{ (PE,ED,CO) }\\[7pt]
    H_{0j}(t_x)z_{f_j}+\big\{H_{0j}(t)-H_{0j}(t_x)\big\}z_{f_j} \mathrm{exp}\big (\beta_j \big )                                            & \text{if } t \geq t_x \text{ (PE) }\\[7pt]
    H_{0j}(t_x)z_{f_j} + \int_{t_x}^t h_{0j}(u)z_{f_j}\mathrm{exp}\big\{ \beta_j  e^{-\eta_j(u-t_x)} \big\} du                     & \text{if } t \geq t_x \text{ (ED) }\\[7pt]
    H_{0j}(t_x)z_{f_j} + \int_{t_x}^t h_{0j}(u)z_{f_j}\mathrm{exp}\big\{ \beta_j  e^{-\eta_j(u-t_x)} + \eta_{0_j} \big\} du           &  \text{if } t \geq t_x \text{ (CO) }
\end{cases}
\end{align*}
where $H_{0j}(t)=\int_{0}^{t} h_{0j}(u) du$ and numerical integration is required for computing cumulative hazard for ED and CO since no closed form exists. 

\newpage
\section*{Web Appendix B: Derivation of cause-specific penetrance function with a time-varying covariate}

Consider a binary time varying covariate $x_{f_i}(t)$ = 0 at $t<t_x$ and 1 at $t \geq t_x$.

\noindent
If $t < t_x$ regardless of TVC models, the marginal cause-specific panetrance function (eq. 6) for event $j$ becomes:
\begin{eqnarray}
F_{f_{ij}}(t|X_{f_i}(t)) & = &
\int_{0}^{t} h_{0j}(u) 
\bigg\{ 1+ \frac{ H_{0j}(u)   }{ \omega_1 } \bigg\}^{-k_j} 
\prod_{l\ne j} \bigg\{1+ \frac{ H_{0l}(u)   }{ \omega_l } \bigg\}^{-k_l} 
\bigg\{1+  \sum_{l=1}^J \frac{ H_{0l}(u)   }{ \omega_l }  \bigg\}^{-k_0}  \nonumber \\
&& \times 
\left [ 
\frac{k_0}{\omega_j} \bigg\{1+ \sum_{l=1}^J \frac{ H_{0l}(u)   }{ \omega_l }  \bigg\}^{-1}  +
\frac{k_j}{\omega_j} \bigg\{ 1+ \frac{ H_{0j}(u)   }{ \omega_j } \bigg\}^{-1} 
\right ]
du,
\nonumber     
\end{eqnarray}  
and if $t \geq t_x$, the cumulative hazard functions $H_{0j}(u)$ will be replaced by $H_{0j}(t_x, u)$ depending on TVC models,
where 
\begin{align*}
H_{0j}(t_x, u)  = 
\begin{cases}
H_{0j}(t_x) + \{ H_{0j}(u)-H_{0j}(t_x)\} \mathrm{exp} ( \beta_j ) & \text{for PE} \\
H_{0j}(t_x) + \int_{t_x}^u h_{0j}(s) \mathrm{ exp}\big\{ \beta_j  e^{-\eta_j(s-t_x)} \big\} ds & \text{for ED} \\
H_{0j}(t_x) + \int_{t_x}^u h_{0j}(s)\mathrm{exp}\big\{ \beta_j  e^{-\eta_j(s-t_x)} + \eta_{0_j} \big\} ds & \text{for CO}
\end{cases}
\end{align*}
Then, the marginal cause-specific panetrance function (eq. 6) for event $j$ for $t>t_x$ can be expressed as:
\begin{eqnarray*}
F_{f_{i1}}(t|X_{f_i}(t))&=&
\int_{0}^{t_x} h_{01}(u) 
\bigg\{ 1+ \frac{ H_{0j}(u)   }{ \omega_1 } \bigg\}^{-k_j} 
\prod_{l\ne j} \bigg\{1+ \frac{ H_{0l}(u)   }{ \omega_l } \bigg\}^{-k_l} 
\bigg\{1+  \sum_{l=1}^J \frac{ H_{0l}(u)   }{ \omega_l }  \bigg\}^{-k_0}  \nonumber \\
&& \times 
\left [ 
\frac{k_0}{\omega_j} \bigg\{1+ \sum_{l=1}^J \frac{ H_{0l}(u)   }{ \omega_l }  \bigg\}^{-1}  +
\frac{k_j}{\omega_j} \bigg\{ 1+ \frac{ H_{0j}(u)   }{ \omega_j } \bigg\}^{-1} 
\right ]
du, \nonumber \\
&& + \int_{t_x}^{t} h_{0j}(u) \exp(\beta_j)
\bigg \{ 1+ \frac{ H_{0j}(t_x, u)  }{ \omega_j } \bigg \}^{-k_j}  \prod_{l\ne j} \bigg \{1+ \frac{ H_{0l}(t_x, u) }{ \omega_l} \bigg \}^{-k_l} \nonumber \\
&& \times \bigg\{1+ \sum_{l=1}^J \frac{ H_{0l}(t_x, u) }{ \omega_l } \bigg\}^{-k_0}  \nonumber \\
&& \times 
\left [ 
\frac{k_0}{\omega_j} \bigg\{1+ \sum_{l=1}^J \frac{ H_{0l}(t_x, u) }{ \omega_l } \bigg\}^{-1}  +
\frac{k_j}{\omega_j} \bigg\{ 1+ \frac{ H_{0j}(t_x, u)  }{ \omega_j } \bigg\}^{-1} 
\right ]
du. \nonumber \\    
\end{eqnarray*}

In the case of independent frailties ($k_0=0$), the marginal cause-specific panetrance function can be simplified as\\
for $t<t_x$
\begin{eqnarray}
F_{f_{ij}}(t|X_{f_i}(t))=
\int_{0}^{t} h_{0j}(u) 
\bigg\{ 1+ \frac{ H_{01}(u)   }{ k_j } \bigg\}^{-k_j-1} 
\prod_{l\ne j} \bigg\{1+ \frac{ H_{0l}(u)   }{ k_l } \bigg\}^{-k_l} du,
\nonumber     
\end{eqnarray}  
and for $t \geq t_x$, 
\begin{eqnarray}
F_{f_{i1}}(t|X_{f_i}(t)) &= & 
\int_{0}^{t_x} h_{0j}(u) 
\bigg\{ 1+ \frac{ H_{01}(u)   }{ k_j } \bigg\}^{-k_j-1} 
\prod_{l\ne j} \bigg\{1+ \frac{ H_{0l}(u)   }{ k_l } \bigg\}^{-k_l} du \nonumber \\
& & + 
\int_{t_x}^{t} h_{0j}(u) \exp(\beta_j) 
\bigg\{ 1+ \frac{ H_{0j}(t_x, u)   }{ k_j } \bigg\}^{-k_j-1} 
\prod_{l\ne j} \bigg\{1+ \frac{ H_{0l}(t_x, u)   }{ k_l } \bigg\}^{-k_l} du \nonumber     
\end{eqnarray}

\newpage
\section*{Web Appendix C: Detailed simulation process}

Data were simulated with code modified from the R package {\tt \bf FamEvent} (Choi et al., 2017). Generation of the cause-specific competing risks survival data is based on the algorithm proposed by Beyersmann et al. (2009). Data generation and analyses were performed using R version 3.4.3.

We consider the shared frailty competing risk model with a TVC and two competing risks. For the covariates, we include one TIC and one TVC.

\begin{enumerate}
\item $G$: Binary mutation status TIC. If the individual is a mutation carrier, G takes value of 1 otherwise 0. We assume cause specific hazards for both competing events are affected by this variable.

\item $x(t)$:  Binary TVC at time $t$. $x(t)=1$ if $t \geq t_s$ and 0 otherwise, where $t_s$ is the time that changes in value of covariate occurred. We assume only the cause-specific hazard for event 1 is affected by this variable.  
\end{enumerate}

The cause-specfic hazards functions for event 1 and event 2 are respectively as follow:
\begin{eqnarray}
h_1(t|X(t),G,z_1)&=&h_{01}(t) \mathrm{exp}\{ \beta_{1gene}G + \mu(x(t)) \}z_{1} \nonumber \\ 
h_2(t|G,z_2)&=&h_{02}(t) \mathrm{exp}\{ \beta_{2gene}G \} z_{2}, \label{eqn:simmodel}
\end{eqnarray}
where $h_{01}(t)$ and $h_{02}(t)$ are the Weibull baseline hazard functions, $z_{1}$ and $z_{2}$ are the cause-specfic shared frailties, $\beta_{1gene}$ and $\beta_{2gene}$ are the mutation status covariate coefficients for event 1 and 2, respectively, and $\mu(x(t))$ is the effect of the binary TVC, which takes the following form depending on the model:    
$$
\mu(x(t))=
\begin{cases}
    0                                                       & \quad \text{if } t < t_s \text{ (PE,ED,CO) }\\
    \beta_{tvc}                                                       & \quad \text{if } t \geq t_s \text{ (PE) }\\
    \beta_{tvc}\, \mathrm{exp}\big\{-\eta(t-t_s)\big\}                     & \quad \text{if } t \geq t_s \text{ (ED) }\\
    \beta_{tvc}\, \mathrm{exp}\big\{-\eta(t-t_s)\big\} + \eta_0          & \quad \text{if } t \geq t_s \text{ (CO) }.
\end{cases}
$$

The algorithm for generating families takes the following three steps based on model (\ref{eqn:simmodel}). Parameters specified in the data generation process, such as the number of siblings for each generation in family pedigree and the current age distribution of the probands and other family members result in the family structure similar to the real data in the application section.\\

Step 1: Family structure
\begin{enumerate}
\item For each family, we generate a three-generation pedigree. We fix two members in the first generation while we generate 2 to 5 siblings in the second and 0 to 2 siblings in the third generations from a truncated negative binomial distribution. 

\item Generate the current age of the proband, ${a_{f_p}}$ from normal distribution with mean age of 45 and SD of 10. Then we generate the current ages of other family members, $\{a_{f_2}, ..., a_{f_i}\}$ for individual $i$, $i=2,\ldots,n_f$, from a normal distribution. The current ages of the first generation are generated with the mean age equal to ${a_{f_p}}+20$ with SD of 1.5 years. The current ages of the second generation are generated from mean age equivalent to ${a_{f_p}}$ with SD of 1.5 years. Finally, for the third generation, their current ages are generated with the mean age subtracted by 20 years from the minimum age of their parents.

\item To generate the TVC status, we first generate $t_{s}$, the time that the TVC occurs,  for all members of the family from a normal distribution with mean age of 40 and variance of 2 years.  If $t_{s,f_i} > a_{f_i}$, we assume no TVC occurred for this individual before their age.

\item Generate the shared frailties $\boldsymbol{z}_f =\{z_{f_1},z_{f_2}\}$ for family $f$ for two competing events. We assume $z_{f_1}$ and $z_{f_2}$ are independent and marginally follow the gamma distribution with shape parameter $k_{1}$ and the scale parameter $1/k_{1}$ for event 1 and $k_{2}$ and $1/k_{2}$ for event 2, respectively. 

\item Generate the mutation status variable $G_{f_p}$ for the proband assuming all the probands are the mutation carriers, based on a dominant model with prespecified \textit{BRCA1} mutation allele frequency of 0.0021. Other family members' mutation statuses are generated conditioning on the proband's mutation status from a Bernoulli distribution with a probability of success equal to $P(G_{f_i}=1|G_{f_p})$. This probability depends only on the relationship between the proband and the $i$th member of the family by Mendelian inheritance laws.

\end{enumerate}

Step 2: Event times and event types
\begin{enumerate}

\item Generate $t_{f_i}$ from the overall survival function: Generate $w$ following a
uniform on $[0, 1]$ and solve for $t_{f_i}$ from $P(T_{f_i} > t_{f_i}| G_{f_i}, t_{s,f_i}, \boldsymbol{z}_{f}) = w$.

\item Given $t_{f_i}$, we decide the event type $\delta_{f_i}$ among two competing events using the rate of the cause-specific hazards at $t_{f_i}$. Compute $h_{1}(t_{f_i}| G_{f_i}, t_{s,f_i}, \boldsymbol{z}_{f})$, $h_{2}(t_{f_i}| G_{f_i}, t_{s,f_i}, \boldsymbol{z}_{f})$ and $p=\frac{h_{1}}{h_{1}+h_{2}}$. Run a Bernoulli experiment with the probability of success $p$. If success, then $\delta_{f_i}=1$ otherwise $\delta_{f_i}=2$. If $t_{f_i} > a_{f_i}$ we regard this individual as censored and $\delta_{f_i}=0$. Follow-up duration is defined from age 16 to $a_{f_i}$ if the individual is right censored, otherwise it is from age 16 to $t_{f_i}$.

\end{enumerate}

Step 3: Ascertainment condition for the family
\begin{enumerate}

\item After generating the event times and types of the family members, keep the family if it satisfies the condition $t_{f_p} < a_{f_p}$. This condition mimics the population based design of the family studies (Gong and Whittemore, 2003) where probands are affected before their study entry age, $a_{f_p}$.

\item Remove men in the pedigree since the real data only consists of women. Mean pedigree size of 5 leads to the total number of individuals about 2500 when 500 families are generated, which agrees with \textit{BRCA1} data.
\end{enumerate}

To generate the data, we specify the following parameters:

\begin{enumerate}
\item baseline hazard function parameters: $\lambda_{1}$ and $\rho_{1}$ for event 1, $\lambda_{2}$ and $\rho_{2}$ for event 2
\item parameters involved in TIC: $\beta_{1gene}$ and $\beta_{2gene}$ as genetic effects for each event
\item parameters involved in TVC:  $\beta_{tvc}$ as a TVC effect for event 1 at the time of TVC occurrence, $\eta$ for ED and CO, additional $\eta_{0}$ for CO
\item familial dependence parameter: $k_1$ and $k_2$ for each event
\end{enumerate}

\newpage
\section*{Web Appendix D: Martingale residuals}

We evaluate the goodness-of-fit using martingale residuals, which are defined as the difference between the number of events of subject $i$ in family $f$ until time $t$ and the expected number of events computed by the cumulative hazard by time $t$. For clustered family data, we define the martingale residuals at both individual level and family level.\\

\noindent
{\bf Martingale residuals at individual level:}\\
The individual martingale residuals for each competing event $j$ are calculated at $T_{f_ij}$, that is at the end of the follow-up as 
$$M_{fij}=  I(\delta_{f_i}=j) - \int_0^{T_{fij}} h_{fij}(t | X_{fi}(t), \hat z_{fj}, \hat\theta ) dt, \mbox{~~for~~} j=1, \ldots, J \, . $$
where $I(\delta_{f_i}=j)$ indicates the occurrence of event $j$.

For the calculation of the cumulated hazards we use the parameters estimated in the model, $\hat\theta$, and the frailties, $\hat z_{fj}$, 
estimated by a posterior distribution of the frailties given the observed data over time. The $\hat z_{fj}$ is obtained by the posterior expectation of the frailties:
$$\hat z_{fj} = E(z_{f j} | T_{f_i j}, X_{f_i}(T_{f_i j}), \hat\theta ) = \frac{ d_{f_j} + \hat\kappa_j + \hat\kappa_0 }{ H_{f_{ij}} (T_{f_i j}| X_{f_i}(T_{f_i j}), \hat\theta) + \hat\kappa_j + \hat\kappa_0 }, $$
where 
$X_{f_i}(T_{f_i j}) $ is the covariate history up to $T_{f_i j}$, the end of the follow-up, $d_{f_j}$ is the number of event $j$ observed in family $f$, 
$H_{f_{ij}} (T_{f_i j}| X_{f_i}(T_{f_i j}), \hat\theta)$ is the cumulative hazard estimated by $T_{f_i j}$, and $\hat\theta$, $\hat\kappa_0$ and $\hat\kappa_j$ are the parameters estimated in the model.  \\

\noindent
{\bf Martingale residuals at family level:}\\
The family level martingale residuals are obtained as the mean martingale residuals by aggregating the individual level martingale residuals in each family
 $$\bar M_{fj} =  \sum_{i=1}^{n_f} M_{fij} / n_f, $$ 
 where $n_f$ is the size of family $f$.
 
The assessment of the model can be performed visually. The mean of the martingale residuals is expected to be equal to 0.


\renewcommand{\baselinestretch}{1}

\begin{table}
\caption{Empirical parameter estimates for the competing risks model with a time varying covariate (TVC) under low ($k_1=7$), medium ($k_1=3.5$) and high ($k_1=1$) familial dependence;  permanent exposure (PE), exponential decay (ED) or Cox and Oaks (CO) models are considered for TVC. For each scenario, the mean bias, empirical standard error (ESE), average standard error (ASE) and estimated 95\% coverage probability (ECP) are obtained from 500 replicates each with on $n=1000$ families.
}

\begin{adjustbox}{width=1\textwidth,center=\textwidth}
{\begin{tabular}{llrrrrrlrrrrrlrrrrr}\\
\hline
\multicolumn{1}{l}{TVC} & \multicolumn{1}{c}{} & \multicolumn{1}{c}{True} & \multicolumn{4}{c}{$k_1=7$, $\tau=0.07$} & \multicolumn{1}{c}{} & \multicolumn{1}{c}{True} & \multicolumn{4}{c}{$k_1=3.5$, $\tau=0.13$} & \multicolumn{1}{c}{} & \multicolumn{1}{c}{True} & \multicolumn{4}{c}{$k_1=1$, $\tau=0.33$} \\ \cline{4-7} \cline{10-13} \cline{16-19} 
\multicolumn{1}{l}{model} & \multicolumn{1}{c}{} & \multicolumn{1}{c}{value} & \multicolumn{1}{c}{Bias} & \multicolumn{1}{c}{ESE} & \multicolumn{1}{c}{ASE} & \multicolumn{1}{c}{ECP} & \multicolumn{1}{c}{} & \multicolumn{1}{c}{value} & \multicolumn{1}{c}{Bias} & \multicolumn{1}{c}{ESE} & \multicolumn{1}{c}{ASE} & \multicolumn{1}{c}{ECP} & \multicolumn{1}{c}{} & \multicolumn{1}{c}{value} & \multicolumn{1}{c}{Bias} & \multicolumn{1}{c}{ESE} & \multicolumn{1}{c}{ASE} & \multicolumn{1}{c}{ECP} \\ \hline
PE & $\mathrm{log}(\lambda_1)$ & -4.83 & -0.01 & 0.04 & 0.04 & 0.95 &  & -4.83 & 0.00 & 0.04 & 0.04 & 0.95 &  & -4.83 & 0.00 & 0.04 & 0.04 & 0.96 \\
 & $\mathrm{log}(\rho_1)$ & 0.88 & 0.00 & 0.02 & 0.02 & 0.96 &  & 0.88 & 0.00 & 0.02 & 0.02 & 0.95 &  & 0.88 & 0.00 & 0.02 & 0.02 & 0.97 \\
 & $\mathrm{log}(\lambda_2)$ & -4.96 & 0.00 & 0.06 & 0.07 & 0.96 &  & -4.96 & -0.01 & 0.07 & 0.07 & 0.96 &  & -4.96 & 0.00 & 0.07 & 0.07 & 0.96 \\
 & $\mathrm{log}(\rho_2)$ & 1.12 & 0.00 & 0.05 & 0.05 & 0.95 &  & 1.12 & 0.00 & 0.05 & 0.05 & 0.97 &  & 1.12 & 0.00 & 0.05 & 0.05 & 0.95 \\
 & $\beta_{1gene}$ & 1.95 & 0.01 & 0.08 & 0.08 & 0.94 &  & 1.95 & 0.00 & 0.08 & 0.08 & 0.94 &  & 1.95 & 0.00 & 0.08 & 0.08 & 0.94 \\
 & $\beta_{2gene}$ & 1.19 & 0.00 & 0.15 & 0.16 & 0.97 &  & 1.19 & 0.01 & 0.16 & 0.16 & 0.95 &  & 1.19 & 0.01 & 0.16 & 0.17 & 0.96 \\
 & $\beta_{1tvc}$ & 0.67 & 0.00 & 0.08 & 0.08 & 0.96 &  & 0.67 & 0.00 & 0.08 & 0.08 & 0.94 &  & 0.67 & 0.00 & 0.08 & 0.08 & 0.94 \\
 & $\mathrm{log}(k_1)$ & 1.95 & 0.12 & 0.62 & 0.58 & 0.95 &  & 1.25 & 0.05 & 0.36 & 0.32 & 0.96 &  & 0.00 & 0.01 & 0.17 & 0.18 & 0.96 \\
 & $\mathrm{log}(k_2)$ & 1.06 & 0.48 & 1.95 & 0.98 & 0.88 &  & 1.06 & 0.38 & 1.38 & 1.01 & 0.91 &  & 1.06 & 0.54 & 2.15 & 1.00 & 0.88 \\
 &  &  &  &  &  &  &  &  &  &  &  &  &  &  &  &  &  &  \\
ED & $\mathrm{log}(\lambda_1)$ & -4.83 & 0.00 & 0.04 & 0.04 & 0.95 &  & -4.83 & 0.00 & 0.04 & 0.04 & 0.95 &  & -4.83 & 0.00 & 0.04 & 0.04 & 0.95 \\
 & $\mathrm{log}(\rho_1)$ & 0.83 & 0.00 & 0.02 & 0.02 & 0.95 &  & 0.83 & 0.00 & 0.02 & 0.02 & 0.95 &  & 0.83 & 0.00 & 0.02 & 0.02 & 0.95 \\
 & $\mathrm{log}(\lambda_2)$ & -4.96 & -0.01 & 0.06 & 0.07 & 0.96 &  & -4.96 & 0.00 & 0.06 & 0.07 & 0.96 &  & -4.96 & 0.00 & 0.06 & 0.07 & 0.98 \\
 & $\mathrm{log}(\rho_2)$ & 1.08 & 0.00 & 0.04 & 0.05 & 0.96 &  & 1.08 & 0.00 & 0.04 & 0.05 & 0.96 &  & 1.08 & 0.00 & 0.04 & 0.04 & 0.96 \\
 & $\beta_{1gene}$ & 1.86 & 0.00 & 0.08 & 0.08 & 0.96 &  & 1.86 & 0.00 & 0.08 & 0.08 & 0.96 &  & 1.86 & 0.01 & 0.08 & 0.08 & 0.96 \\
 & $\beta_{2gene}$ & 1.22 & 0.00 & 0.15 & 0.15 & 0.94 &  & 1.22 & 0.02 & 0.14 & 0.15 & 0.97 &  & 1.22 & 0.01 & 0.15 & 0.15 & 0.96 \\
 & $\beta_{1tvc}$ & 1.87 & 0.01 & 0.18 & 0.18 & 0.94 &  & 1.87 & 0.01 & 0.16 & 0.18 & 0.96 &  & 1.87 & 0.02 & 0.17 & 0.17 & 0.95 \\
 & $\mathrm{log}(\eta)$ & -1.28 & 0.01 & 0.21 & 0.21 & 0.95 &  & -1.28 & 0.00 & 0.20 & 0.22 & 0.96 &  & -1.28 & 0.03 & 0.20 & 0.21 & 0.96 \\
 & $\mathrm{log}(k_1)$ & 1.95 & 0.16 & 0.91 & 0.59 & 0.94 &  & 1.25 & 0.04 & 0.32 & 0.32 & 0.95 &  & 0.00 & 0.02 & 0.16 & 0.17 & 0.97 \\
 & $\mathrm{log}(k_2)$ & 1.18 & 0.34 & 1.18 & 0.92 & 0.91 &  & 1.18 & 0.36 & 1.58 & 0.97 & 0.90 &  & 1.18 & 0.41 & 1.37 & 1.04 & 0.92 \\
 &  &  &  &  &  &  &  &  &  &  &  &  &  &  &  &  &  &  \\
CO & $\mathrm{log}(\lambda_1)$ & -4.83 & 0.00 & 0.03 & 0.04 & 0.95 &  & -4.83 & 0.00 & 0.03 & 0.04 & 0.98 &  & -4.83 & 0.00 & 0.03 & 0.04 & 0.96 \\
 & $\mathrm{log}(\rho_1)$ & 0.83 & 0.00 & 0.02 & 0.02 & 0.95 &  & 0.83 & 0.00 & 0.02 & 0.02 & 0.98 &  & 0.83 & 0.00 & 0.02 & 0.02 & 0.96 \\
 & $\mathrm{log}(\lambda_2)$ & -4.96 & 0.00 & 0.05 & 0.06 & 0.95 &  & -4.96 & -0.01 & 0.05 & 0.06 & 0.96 &  & -4.96 & 0.00 & 0.05 & 0.06 & 0.97 \\
 & $\mathrm{log}(\rho_2)$ & 1.07 & 0.00 & 0.03 & 0.04 & 0.97 &  & 1.07 & 0.00 & 0.03 & 0.04 & 0.97 &  & 1.07 & 0.00 & 0.03 & 0.04 & 0.97 \\
 & $\beta_{1gene}$ & 2.08 & 0.01 & 0.07 & 0.08 & 0.97 &  & 2.08 & 0.01 & 0.07 & 0.08 & 0.95 &  & 2.08 & 0.00 & 0.07 & 0.08 & 0.95 \\
 & $\beta_{2gene}$ & 1.57 & 0.00 & 0.12 & 0.14 & 0.95 &  & 1.57 & 0.01 & 0.11 & 0.15 & 0.97 &  & 1.57 & 0.01 & 0.12 & 0.15 & 0.96 \\
 & $\beta_{1tvc}$ & 1.52 & 0.02 & 0.26 & 0.30 & 0.93 &  & 1.52 & 0.04 & 0.24 & 0.31 & 0.96 &  & 1.52 & 0.04 & 0.23 & 0.30 & 0.96 \\
 & $\mathrm{log}(\eta)$ & -0.18 & 0.02 & 0.35 & 0.42 & 0.93 &  & -0.18 & 0.03 & 0.33 & 0.43 & 0.94 &  & -0.18 & 0.04 & 0.36 & 0.43 & 0.94 \\
 & $\eta_0$ & 0.21 & 0.00 & 0.08 & 0.09 & 0.96 &  & 0.21 & 0.00 & 0.08 & 0.10 & 0.97 &  & 0.21 & -0.01 & 0.08 & 0.10 & 0.96 \\ 
 & $\mathrm{log}(k_1)$ & 1.95 & 0.03 & 0.43 & 0.51 & 0.94 &  & 1.25 & 0.04 & 0.25 & 0.30 & 0.97 &  & 0.00 & 0.00 & 0.13 & 0.16 & 0.97 \\
 & $\mathrm{log}(k_2)$ & 1.26 & 0.27 & 0.72 & 0.92 & 0.93 &  & 1.26 & 0.14 & 0.71 & 0.83 & 0.92 &  & 1.26 & 0.19 & 0.68 & 0.87 & 0.92 \\ \hline
\multicolumn{19}{l}{$\lambda_j$ and $\rho_j$ are baseline hazard parameters for event $j, j=1,2$; $\beta_{jgene}$ is the coefficient of a time-invariant covariate for event $j$;}\\
\multicolumn{19}{l}{$\beta_{1tvc}$, $\eta$ and $\eta_0$ are parameters to describe TVC effects; $k_j$ is the frailty parameter for event $j$.  }\\ 
\end{tabular}}
\end{adjustbox}
\bigskip
\end{table}

\begin{table}
\caption{Empirical penetrance estimates by age 70 for the competing risks model with a time varying covariate (TVC) under low ($k_1=7$), medium ($k_1=3.5$) and high ($k_1=1$) familial dependence;  permanent exposure (PE), exponential decay (ED) or Cox and Oaks (CO) models are considered for TVC; $F_1(70; S, G)$ and $F_2(70; S, G)$ are cause-specific penetrance estimators (\%) by age 70 for event 1 and event 2, respectively, given TVC and mutation status (G), and TVC occurred at age 35 if $S=1$. For each scenario, the mean bias, empirical standard error (ESE), average standard error (ASE) and estimated 95\% coverage probability (ECP) are obtained from 500 replicates each $n=1000$ families. 
}
\label{wt:two}
\begin{adjustbox}{width=1\textwidth,center=\textwidth}
{\begin{tabular}{llrrrrrlrrrrrlrrrrr}\\
\hline
\multicolumn{1}{l}{TVC} & \multicolumn{1}{c}{} & \multicolumn{1}{c}{True} & \multicolumn{4}{c}{$k_1=7$, $\tau=0.07$} & \multicolumn{1}{c}{} & \multicolumn{1}{c}{True} & \multicolumn{4}{c}{$k_1=3.5$, $\tau=0.13$} & \multicolumn{1}{c}{} & \multicolumn{1}{c}{True} & \multicolumn{4}{c}{$k_1=1$, $\tau=0.33$} \\ \cline{4-7} \cline{10-13} \cline{16-19} 
\multicolumn{1}{l}{model} & \multicolumn{1}{c}{} & \multicolumn{1}{c}{value} & \multicolumn{1}{c}{Bias} & \multicolumn{1}{c}{ESE} & \multicolumn{1}{c}{ASE} & \multicolumn{1}{c}{ECP} & \multicolumn{1}{c}{} & \multicolumn{1}{c}{value} & \multicolumn{1}{c}{Bias} & \multicolumn{1}{c}{ESE} & \multicolumn{1}{c}{ASE} & \multicolumn{1}{c}{ECP} & \multicolumn{1}{c}{} & \multicolumn{1}{c}{value} & \multicolumn{1}{c}{Bias} & \multicolumn{1}{c}{ESE} & \multicolumn{1}{c}{ASE} & \multicolumn{1}{c}{ECP} \\ \hline
PE & $F_1(70; \mbox{TVC = 0, G = 0})$ & 12.56 & -0.09 & 0.99 & 0.97 & 0.94 &  & 12.45 & -0.01 & 1.02 & 0.98 & 0.95 &  & 11.93 & -0.02 & 0.99 & 1.02 & 0.95 \\
 & $F_1(70; \mbox{TVC = 1, G = 0})$ & 21.92 & -0.08 & 1.78 & 1.74 & 0.95 &  & 21.58 & -0.03 & 1.79 & 1.76 & 0.94 &  & 20.09 & -0.04 & 1.73 & 1.76 & 0.95 \\
 & $F_1(70; \mbox{TVC = 0, G = 1})$ & 56.52 & -0.09 & 2.19 & 2.27 & 0.96 &  & 54.51 & 0.03 & 2.41 & 2.42 & 0.96 &  & 46.80 & 0.03 & 2.67 & 2.77 & 0.96 \\
 & $F_1(70; \mbox{TVC = 1, G = 1})$ & 75.63 & -0.04 & 2.57 & 2.66 & 0.95 &  & 72.59 & -0.06 & 2.84 & 2.88 & 0.95 &  & 61.08 & 0.03 & 3.27 & 3.38 & 0.96 \\
 &  &  &  &  &  &  &  &  &  &  &  &  &  &  &  &  &  &  \\
 & $F_2(70; \mbox{TVC = 0, G = 0})$ & 4.73 & 0.03 & 0.58 & 0.61 & 0.96 &  & 4.73 & -0.02 & 0.60 & 0.61 & 0.96 &  & 4.74 & 0.00 & 0.61 & 0.62 & 0.94 \\
 & $F_2(70; \mbox{TVC = 1, G = 0})$ & 4.45 & 0.02 & 0.55 & 0.57 & 0.95 &  & 4.45 & -0.02 & 0.57 & 0.57 & 0.95 &  & 4.49 & 0.00 & 0.58 & 0.59 & 0.94 \\
 & $F_2(70; \mbox{TVC = 0, G = 1})$ & 9.68 & 0.02 & 0.80 & 0.82 & 0.95 &  & 9.85 & 0.02 & 0.77 & 0.83 & 0.97 &  & 10.52 & 0.01 & 0.88 & 0.91 & 0.96 \\
 & $F_2(70; \mbox{TVC = 1, G = 1})$ & 7.12 & 0.00 & 0.61 & 0.63 & 0.96 &  & 7.42 & 0.02 & 0.62 & 0.65 & 0.96 &  & 8.56 & 0.00 & 0.72 & 0.73 & 0.95 \\
 &  &  &  &  &  &  &  &  &  &  &  &  &  &  &  &  &  &  \\
ED & $F_1(70; \mbox{TVC = 0, G = 0})$ & 13.55 & 0.01 & 0.99 & 1.01 & 0.96 &  & 13.42 & 0.01 & 0.96 & 1.02 & 0.95 &  & 12.82 & 0.05 & 1.02 & 1.04 & 0.95 \\
 & $F_1(70; \mbox{TVC = 1, G = 0})$ & 15.49 & 0.03 & 1.09 & 1.16 & 0.96 &  & 15.32 & 0.06 & 1.13 & 1.17 & 0.95 &  & 14.54 & 0.04 & 1.15 & 1.18 & 0.95 \\
 & $F_1(70; \mbox{TVC = 0, G = 1})$ & 55.65 & -0.10 & 2.03 & 2.19 & 0.98 &  & 53.68 & -0.08 & 2.20 & 2.31 & 0.96 &  & 46.14 & 0.27 & 2.53 & 2.60 & 0.95 \\
 & $F_1(70; \mbox{TVC = 1, G = 1})$ & 60.49 & -0.05 & 2.15 & 2.36 & 0.97 &  & 58.24 & 0.02 & 2.42 & 2.49 & 0.95 &  & 49.69 & 0.25 & 2.68 & 2.78 & 0.95 \\
 &  &  &  &  &  &  &  &  &  &  &  &  &  &  &  &  &  &  \\
 & $F_2(70; \mbox{TVC = 0, G = 0})$ & 5.39 & -0.03 & 0.60 & 0.64 & 0.95 &  & 5.39 & -0.05 & 0.61 & 0.65 & 0.96 &  & 5.41 & -0.01 & 0.62 & 0.66 & 0.96 \\
 & $F_2(70; \mbox{TVC = 1, G = 0})$ & 5.26 & -0.03 & 0.58 & 0.63 & 0.95 &  & 5.26 & -0.05 & 0.60 & 0.63 & 0.95 &  & 5.28 & -0.01 & 0.60 & 0.65 & 0.96 \\
 & $F_2(70; \mbox{TVC = 0, G = 1})$ & 11.38 & -0.04 & 0.82 & 0.86 & 0.96 &  & 11.57 & 0.04 & 0.86 & 0.88 & 0.96 &  & 12.34 & 0.03 & 0.98 & 0.96 & 0.94 \\
 & $F_2(70; \mbox{TVC = 1, G = 1})$ & 9.97 & -0.04 & 0.74 & 0.76 & 0.96 &  & 10.22 & 0.02 & 0.77 & 0.79 & 0.95 &  & 11.20 & 0.03 & 0.88 & 0.87 & 0.94 \\
 &  &  &  &  &  &  &  &  &  &  &  &  &  &  &  &  &  &  \\
CO & $F_1(70; \mbox{TVC = 0, G = 0})$ & 13.54 & -0.10 & 0.99 & 1.00 & 0.95 &  & 13.41 & -0.04 & 0.93 & 1.01 & 0.97 &  & 12.81 & 0.01 & 0.98 & 1.01 & 0.96 \\
 & $F_1(70; \mbox{TVC = 1, G = 0})$ & 16.60 & -0.11 & 1.40 & 1.43 & 0.94 &  & 16.41 & -0.06 & 1.31 & 1.43 & 0.95 &  & 15.52 & -0.02 & 1.33 & 1.40 & 0.96 \\
 & $F_1(70; \mbox{TVC = 0, G = 1})$ & 61.12 & -0.10 & 1.93 & 2.09 & 0.96 &  & 58.82 & 0.18 & 2.11 & 2.19 & 0.95 &  & 50.11 & 0.05 & 2.38 & 2.46 & 0.95 \\
 & $F_1(70; \mbox{TVC = 1, G = 1})$ & 67.55 & -0.13 & 2.58 & 2.65 & 0.95 &  & 64.90 & 0.11 & 2.60 & 2.73 & 0.95 &  & 54.88 & -0.04 & 2.69 & 2.88 & 0.96 \\
 &  &  &  &  &  &  &  &  &  &  &  &  &  &  &  &  &  &  \\
 & $F_2(70; \mbox{TVC = 0, G = 0})$ & 5.53 & 0.02 & 0.64 & 0.65 & 0.95 &  & 5.53 & -0.06 & 0.59 & 0.65 & 0.96 &  & 5.55 & 0.00 & 0.63 & 0.67 & 0.97 \\
 & $F_2(70; \mbox{TVC = 1, G = 0})$ & 5.39 & 0.02 & 0.62 & 0.64 & 0.95 &  & 5.39 & -0.06 & 0.58 & 0.63 & 0.96 &  & 5.42 & 0.00 & 0.61 & 0.65 & 0.97 \\
 & $F_2(70; \mbox{TVC = 0, G = 1})$ & 14.27 & 0.01 & 0.97 & 0.97 & 0.96 &  & 14.61 & -0.12 & 0.93 & 0.99 & 0.97 &  & 15.91 & 0.02 & 1.00 & 1.10 & 0.97 \\
 & $F_2(70; \mbox{TVC = 1, G = 1})$ & 12.35 & 0.01 & 0.91 & 0.91 & 0.94 &  & 12.77 & -0.11 & 0.90 & 0.93 & 0.94 &  & 14.36 & 0.02 & 0.94 & 1.02 & 0.97 \\ \hline
\end{tabular}}
\end{adjustbox}
\bigskip
\end{table}

\begin{table}
\caption{Simulation results under a misspecified time varying covariate (TVC): parameter estimates for the competing risks model with a TVC under low ($k_1=7$), medium ($k_1=3.5$) and high ($k_1=1$) familial dependence;  permanent exposure (PE), exponential decay (ED) or Cox and Oaks (CO) models are considered as TVC. For each scenario, the three TVC models are fitted and the mean bias, empirical standard error (ESE), average standard error (ASE) and estimated 95\% coverage probability (ECP) are obtained from 500 replicates each with $n=500$ families. }\label{tab:mis1}

\begin{adjustbox}{width=1\textwidth,center=\textwidth}
{\begin{tabular}{llrrrrrlrrrrrlrrrrr}\\
\hline
\multicolumn{1}{c}{} & \multicolumn{1}{c}{} & \multicolumn{5}{c}{True Model (PE)} & \multicolumn{1}{c}{} & \multicolumn{5}{c}{Misspecified Model (ED)} & \multicolumn{1}{c}{} & \multicolumn{5}{c}{Misspecified Model (CO)} \\ \cline{3-7} \cline{9-13} \cline{15-19} 
\multicolumn{1}{c}{True} & \multicolumn{1}{c}{} & \multicolumn{1}{c}{True} & \multicolumn{4}{c}{$k_1=3.5$, $\tau=0.13$} & \multicolumn{1}{c}{} & \multicolumn{1}{c}{True} & \multicolumn{4}{c}{$k_1=3.5$, $\tau=0.13$} & \multicolumn{1}{c}{} & \multicolumn{1}{c}{True} & \multicolumn{4}{c}{$k_1=3.5$, $\tau=0.13$} \\ \cline{4-7} \cline{10-13} \cline{16-19} 
\multicolumn{1}{c}{TVC} & \multicolumn{1}{c}{} & \multicolumn{1}{c}{value} & \multicolumn{1}{c}{Bias} & \multicolumn{1}{c}{ESE} & \multicolumn{1}{c}{ASE} & \multicolumn{1}{c}{ECP} & \multicolumn{1}{c}{} & \multicolumn{1}{c}{value} & \multicolumn{1}{c}{Bias} & \multicolumn{1}{c}{ESE} & \multicolumn{1}{c}{ASE} & \multicolumn{1}{c}{ECP} & \multicolumn{1}{c}{} & \multicolumn{1}{c}{value} & \multicolumn{1}{c}{Bias} & \multicolumn{1}{c}{ESE} & \multicolumn{1}{c}{ASE} & \multicolumn{1}{c}{ECP} \\ \hline
PE & $\mathrm{log}(\lambda_1)$ & -4.83 & 0.00 & 0.06 & 0.06 & 0.95 &  & -4.83 & 0.01 & 0.06 & 0.06 & 0.95 &  & -4.83 & 0.00 & 0.06 & 0.06 & 0.95 \\
 & $\mathrm{log}(\rho_1)$ & 0.88 & 0.00 & 0.03 & 0.03 & 0.93 &  & 0.88 & 0.00 & 0.03 & 0.03 & 0.95 &  & 0.88 & 0.00 & 0.03 & 0.03 & 0.95 \\
 & $\mathrm{log}(\lambda_2)$ & -4.96 & -0.02 & 0.10 & 0.10 & 0.94 &  & -4.96 & 0.00 & 0.10 & 0.10 & 0.95 &  & -4.96 & -0.01 & 0.10 & 0.10 & 0.95 \\
 & $\mathrm{log}(\rho_2)$ & 1.12 & 0.00 & 0.07 & 0.07 & 0.95 &  & 1.12 & 0.01 & 0.07 & 0.07 & 0.95 &  & 1.12 & 0.00 & 0.07 & 0.07 & 0.95 \\
 & $\beta_{1tvc}$ & 0.67 & 0.00 & 0.10 & 0.11 & 0.96 &  & 0.67 & 0.05 & 0.13 & 0.13 & 0.93 &  & 0.67 & -0.30 & 1.08 & 0.72 & 0.54 \\
 & $\beta_{1gene}$ & 1.95 & 0.01 & 0.12 & 0.12 & 0.96 &  & 1.95 & 0.00 & 0.12 & 0.12 & 0.95 &  & 1.95 & 0.00 & 0.12 & 0.12 & 0.94 \\
 & $\beta_{2gene}$ & 1.19 & 0.03 & 0.24 & 0.23 & 0.95 &  & 1.19 & 0.01 & 0.23 & 0.23 & 0.96 &  & 1.19 & 0.02 & 0.21 & 0.23 & 0.97 \\
 & $\mathrm{log}(k_1)$ & 1.25 & 0.13 & 0.69 & 0.48 & 0.95 &  & 1.25 & 0.13 & 0.58 & 0.51 & 0.97 &  & 1.25 & 0.08 & 0.54 & 0.50 & 0.96 \\
 & $\mathrm{log}(k_2)$ & 1.06 & 0.72 & 2.20 & 1.41 & 0.84 &  & 1.06 & 0.95 & 2.49 & 1.54 & 0.81 &  & 1.06 & 0.87 & 2.27 & 1.57 & 0.81 \\
 & $\eta$ & - & - & - & - & - &  & 0.00 & 0.01 & 0.01 & 0.01 & 0.93 &  & 0.00 & 0.02 & 0.05 & 0.04 & 0.91 \\
 & $\eta_{0}$ & - & - & - & - & - &  & - & - & - & - & - &  & 0.00 & 0.33 & 1.04 & 0.72 & 0.59 \\
 &  &  &  &  &  &  &  &  &  &  &  &  &  &  &  &  &  &  \\
 &  & \multicolumn{5}{c}{True Model (ED)} & \multicolumn{1}{c}{} & \multicolumn{5}{c}{Misspecified Model (PE)} & \multicolumn{1}{c}{} & \multicolumn{5}{c}{Misspecified Model (CO)} \\ \cline{3-7} \cline{9-13} \cline{15-19} 
ED & $\mathrm{log}(\lambda_1)$ & -4.83 & 0.00 & 0.06 & 0.06 & 0.95 &  & -4.83 & -0.05 & 0.07 & 0.06 & 0.86 &  & -4.83 & 0.00 & 0.06 & 0.06 & 0.95 \\
 & $\mathrm{log}(\rho_1)$ & 0.83 & 0.00 & 0.03 & 0.03 & 0.95 &  & 0.83 & -0.02 & 0.03 & 0.03 & 0.87 &  & 0.83 & 0.00 & 0.03 & 0.03 & 0.95 \\
 & $\mathrm{log}(\lambda_2)$ & -4.96 & -0.01 & 0.09 & 0.09 & 0.96 &  & -4.96 & 0.00 & 0.10 & 0.09 & 0.94 &  & -4.96 & -0.01 & 0.08 & 0.09 & 0.96 \\
 & $\mathrm{log}(\rho_2)$ & 1.08 & 0.00 & 0.06 & 0.06 & 0.95 &  & 1.08 & 0.01 & 0.07 & 0.06 & 0.94 &  & 1.08 & 0.00 & 0.06 & 0.06 & 0.95 \\
 & $\beta_{1tvc}$ & 1.87 & -0.01 & 0.25 & 0.25 & 0.95 &  & 1.87 & -1.37 & 0.13 & 0.13 & 0.00 &  & 1.87 & 0.04 & 0.26 & 0.27 & 0.96 \\
 & $\beta_{1gene}$ & 1.86 & 0.01 & 0.11 & 0.12 & 0.95 &  & 1.86 & 0.04 & 0.13 & 0.12 & 0.95 &  & 1.86 & 0.01 & 0.11 & 0.12 & 0.96 \\
 & $\beta_{2gene}$ & 1.22 & 0.03 & 0.22 & 0.21 & 0.96 &  & 1.22 & 0.01 & 0.23 & 0.22 & 0.95 &  & 1.22 & 0.02 & 0.20 & 0.21 & 0.96 \\
 & $\mathrm{log}(k_1)$ & 1.25 & 0.08 & 0.49 & 0.48 & 0.97 &  & 1.25 & 0.21 & 1.08 & 0.60 & 0.86 &  & 1.25 & 0.11 & 0.55 & 0.49 & 0.96 \\
 & $\mathrm{log}(k_2)$ & 1.18 & 0.53 & 1.70 & 1.26 & 0.84 &  & 1.18 & 0.96 & 1.72 & 2.12 & 0.78 &  & 1.18 & 0.61 & 1.70 & 1.46 & 0.84 \\
 & $\eta$ & 0.28 & 0.02 & 0.09 & 0.09 & 0.94 &  & - & - & - & - & - &  & 0.28 & 0.03 & 0.13 & 0.11 & 0.93 \\
 & $\eta_{0}$ & - & - & - & - & - &  & - & - & - & - & - &  & 0.00 & -0.02 & 0.19 & 0.18 & 0.95 \\
 &  &  &  &  &  &  &  &  &  &  &  &  &  &  &  &  &  &  \\
 &  & \multicolumn{5}{c}{True Model (CO)} & \multicolumn{1}{c}{} & \multicolumn{5}{c}{Misspecified Model (PE)} & \multicolumn{1}{c}{} & \multicolumn{5}{c}{Misspecified Model (ED)} \\ \cline{3-7} \cline{9-13} \cline{15-19} 
CO & $\mathrm{log}(\lambda_1)$ & -4.83 & 0.00 & 0.05 & 0.06 & 0.94 &  & -4.83 & -0.02 & 0.06 & 0.06 & 0.94 &  & -4.83 & 0.03 & 0.06 & 0.05 & 0.90 \\
 & $\mathrm{log}(\rho_1)$ & 0.83 & 0.00 & 0.03 & 0.03 & 0.96 &  & 0.83 & -0.01 & 0.03 & 0.03 & 0.93 &  & 0.83 & 0.01 & 0.03 & 0.03 & 0.94 \\
 & $\mathrm{log}(\lambda_2)$ & -4.96 & 0.00 & 0.07 & 0.09 & 0.97 &  & -4.96 & 0.01 & 0.09 & 0.09 & 0.92 &  & -4.96 & -0.01 & 0.09 & 0.09 & 0.95 \\
 & $\mathrm{log}(\rho_2)$ & 1.07 & 0.00 & 0.05 & 0.06 & 0.97 &  & 1.07 & 0.01 & 0.07 & 0.06 & 0.92 &  & 1.07 & 0.00 & 0.06 & 0.06 & 0.94 \\
 & $\beta_{1tvc}$ & 1.52 & 0.04 & 0.33 & 0.42 & 0.94 &  & 1.52 & -1.15 & 0.13 & 0.12 & 0.00 &  & 1.52 & 0.10 & 0.45 & 0.42 & 0.88 \\
 & $\beta_{1gene}$ & 2.08 & 0.01 & 0.10 & 0.12 & 0.95 &  & 2.08 & 0.02 & 0.12 & 0.12 & 0.94 &  & 2.08 & 0.00 & 0.12 & 0.12 & 0.94 \\
 & $\beta_{2gene}$ & 1.57 & 0.00 & 0.17 & 0.21 & 0.94 &  & 1.57 & 0.03 & 0.24 & 0.21 & 0.92 &  & 1.57 & 0.03 & 0.20 & 0.21 & 0.95 \\
 & $\mathrm{log}(k_1)$ & 1.25 & 0.10 & 0.39 & 0.46 & 0.96 &  & 1.25 & 0.20 & 0.82 & 0.55 & 0.92 &  & 1.25 & 0.08 & 0.48 & 0.44 & 0.96 \\
 & $\mathrm{log}(k_2)$ & 1.26 & 0.35 & 0.98 & 1.40 & 0.90 &  & 1.26 & 0.60 & 1.39 & 1.96 & 0.80 &  & 1.26 & 0.52 & 1.74 & 1.38 & 0.86 \\
 & $\eta$ & 0.83 & 0.01 & 0.59 & 0.56 & 0.91 &  & - & - & - & - & - &  & 0.83 & -0.13 & 0.51 & 0.41 & 0.72 \\
 & $\eta_0$ & 0.21 & -0.01 & 0.12 & 0.14 & 0.96 &  & - & - & - & - & - &  & - & - & - & - & - \\ \hline
\multicolumn{19}{l}{$\lambda_j$ and $\rho_j$ are baseline hazard parameters for event $j, j=1,2$; $\beta_{jgene}$ is the coefficient of a time-invariant covariate for event $j$;}\\
\multicolumn{19}{l}{$\beta_{1tvc}$, $\eta$ and $\eta_0$ are parameters to describe TVC effects; $k_j$ is the frailty parameter for event $j$.  }\\ 

\end{tabular}}
\end{adjustbox}
\bigskip
\end{table}

\begin{table}
\caption{Simulation results under a misspecified time varying covariate (TVC): penetrance estimates by age 70 for a competing risks model with a TVC under low ($k_1=7$), medium ($k_1=3.5$) and high ($k_1=1$) familial dependence;  permanent exposure (PE), exponential decay (ED) or Cox and Oaks (CO) models are considered for TVC; $F_1(70; \mbox{TVC, G})$ and $F_2(70; \mbox{TVC, G})$ are cause-specific penetrance estimators (\%) by age 70 for event 1 and event 2, respectively, given TVC and mutation status (G), and TVC occurred at age 35 if $\mbox{TVC}=1$. For each scenario, the three TVC models are fitted and the mean bias, empirical standard error (ESE), average standard error (ASE) and estimated 95\% coverage probability (ECP) are obtained from 500 replicates each with $n=500$ families. 
}\label{tab:mis2}

\begin{adjustbox}{width=1\textwidth,center=\textwidth}
{\begin{tabular}{llrrrrrlrrrrrlrrrrr}\\
\hline
\multicolumn{1}{c}{} & \multicolumn{1}{c}{} & \multicolumn{5}{c}{True Model (PE)} & \multicolumn{1}{c}{} & \multicolumn{5}{c}{Misspecified Model (ED)} & \multicolumn{1}{c}{} & \multicolumn{5}{c}{Misspecified Model (CO)} \\ \cline{3-7} \cline{9-13} \cline{15-19} 
\multicolumn{1}{c}{True} & \multicolumn{1}{c}{} & \multicolumn{1}{c}{True} & \multicolumn{4}{c}{$k_1=3.5$, $\tau=0.13$} & \multicolumn{1}{c}{} & \multicolumn{1}{c}{True} & \multicolumn{4}{c}{$k_1=3.5$, $\tau=0.13$} & \multicolumn{1}{c}{} & \multicolumn{1}{c}{True} & \multicolumn{4}{c}{$k_1=3.5$, $\tau=0.13$} \\ \cline{4-7} \cline{10-13} \cline{16-19} 
\multicolumn{1}{c}{TVC} & \multicolumn{1}{c}{} & \multicolumn{1}{c}{value} & \multicolumn{1}{c}{Bias} & \multicolumn{1}{c}{ESE} & \multicolumn{1}{c}{ASE} & \multicolumn{1}{c}{ECP} & \multicolumn{1}{c}{} & \multicolumn{1}{c}{value} & \multicolumn{1}{c}{Bias} & \multicolumn{1}{c}{ESE} & \multicolumn{1}{c}{ASE} & \multicolumn{1}{c}{ECP} & \multicolumn{1}{c}{} & \multicolumn{1}{c}{value} & \multicolumn{1}{c}{Bias} & \multicolumn{1}{c}{ESE} & \multicolumn{1}{c}{ASE} & \multicolumn{1}{c}{ECP} \\ \hline
PE & $F_1(70; \mbox{TVC = 0, G = 0})$ & 12.45 & 0.01 & 1.33 & 1.40 & 0.94 &  & 12.45 & 0.21 & 1.41 & 1.43 & 0.95 &  & 12.45 & 0.01 & 1.41 & 1.42 & 0.96 \\
 & $F_1(70; \mbox{TVC = 1, G = 0})$ & 21.58 & 0.02 & 2.37 & 2.48 & 0.95 &  & 21.58 & -0.35 & 2.40 & 2.54 & 0.95 &  & 21.58 & -0.14 & 2.56 & 2.59 & 0.95 \\
 & $F_1(70; \mbox{TVC = 0, G = 1})$ & 54.51 & 0.12 & 3.39 & 3.42 & 0.94 &  & 54.51 & 0.39 & 3.30 & 3.49 & 0.96 &  & 54.51 & -0.23 & 3.30 & 3.48 & 0.95 \\
 & $F_1(70; \mbox{TVC = 1, G = 1})$ & 72.59 & 0.03 & 4.08 & 4.06 & 0.94 &  & 72.59 & -0.62 & 3.97 & 4.26 & 0.96 &  & 72.59 & -0.57 & 4.05 & 4.27 & 0.96 \\
 &  &  &  &  &  &  &  &  &  &  &  &  &  &  &  &  &  &  \\
 & $F_2(70; \mbox{TVC = 0, G = 0})$ & 4.73 & -0.08 & 0.87 & 0.85 & 0.93 &  & 4.73 & 0.01 & 0.88 & 0.87 & 0.94 &  & 4.73 & -0.05 & 0.81 & 0.86 & 0.95 \\
 & $F_2(70; \mbox{TVC = 1, G = 0})$ & 4.45 & -0.08 & 0.82 & 0.80 & 0.93 &  & 4.45 & 0.01 & 0.83 & 0.82 & 0.94 &  & 4.45 & -0.05 & 0.77 & 0.81 & 0.95 \\
 & $F_2(70; \mbox{TVC = 0, G = 1})$ & 9.85 & -0.04 & 1.16 & 1.18 & 0.95 &  & 9.85 & -0.04 & 1.13 & 1.18 & 0.94 &  & 9.85 & 0.05 & 1.05 & 1.18 & 0.97 \\
 & $F_2(70; \mbox{TVC = 1, G = 1})$ & 7.42 & -0.04 & 0.91 & 0.92 & 0.95 &  & 7.42 & -0.01 & 0.87 & 0.93 & 0.96 &  & 7.42 & 0.04 & 0.84 & 0.93 & 0.96 \\
 &  &  &  &  &  &  &  &  &  &  &  &  &  &  &  &  &  &  \\
 &  & \multicolumn{5}{c}{True Model (ED)} & \multicolumn{1}{c}{} & \multicolumn{5}{c}{Misspecified Model (PE)} & \multicolumn{1}{c}{} & \multicolumn{5}{c}{Misspecified Model (CO)} \\ \cline{3-7} \cline{9-13} \cline{15-19} 
ED & $F_1(70; \mbox{TVC = 0, G = 0})$ & 13.42 & -0.02 & 1.41 & 1.44 & 0.94 &  & 13.42 & -0.77 & 1.63 & 1.52 & 0.87 &  & 13.42 & -0.02 & 1.39 & 1.49 & 0.96 \\
 & $F_1(70; \mbox{TVC = 1, G = 0})$ & 15.32 & 0.05 & 1.62 & 1.66 & 0.94 &  & 15.32 & 3.61 & 2.38 & 2.29 & 0.71 &  & 15.32 & -0.01 & 1.99 & 2.01 & 0.95 \\
 & $F_1(70; \mbox{TVC = 0, G = 1})$ & 53.68 & -0.05 & 3.03 & 3.27 & 0.96 &  & 53.68 & -1.11 & 3.79 & 3.54 & 0.90 &  & 53.68 & 0.04 & 3.29 & 3.39 & 0.95 \\
 & $F_1(70; \mbox{TVC = 1, G = 1})$ & 58.24 & 0.10 & 3.26 & 3.54 & 0.97 &  & 58.24 & 7.37 & 4.35 & 4.28 & 0.61 &  & 58.24 & -0.03 & 4.20 & 4.36 & 0.96 \\
 &  &  &  &  &  &  &  &  &  &  &  &  &  &  &  &  &  &  \\
 & $F_2(70; \mbox{TVC = 0, G = 0})$ & 5.39 & -0.07 & 0.86 & 0.92 & 0.95 &  & 5.39 & 0.07 & 0.98 & 0.97 & 0.94 &  & 5.39 & -0.05 & 0.85 & 0.92 & 0.95 \\
 & $F_2(70; \mbox{TVC = 1, G = 0})$ & 5.26 & -0.07 & 0.83 & 0.89 & 0.95 &  & 5.26 & -0.03 & 0.94 & 0.93 & 0.94 &  & 5.26 & -0.05 & 0.83 & 0.90 & 0.95 \\
 & $F_2(70; \mbox{TVC = 0, G = 1})$ & 11.57 & 0.04 & 1.29 & 1.24 & 0.95 &  & 11.57 & 0.20 & 1.31 & 1.31 & 0.95 &  & 11.57 & 0.02 & 1.16 & 1.25 & 0.97 \\
 & $F_2(70; \mbox{TVC = 1, G = 1})$ & 10.22 & 0.01 & 1.12 & 1.12 & 0.95 &  & 10.22 & -0.53 & 1.08 & 1.14 & 0.90 &  & 10.22 & 0.01 & 1.08 & 1.16 & 0.96 \\
 &  &  &  &  &  &  &  &  &  &  &  &  &  &  &  &  &  &  \\
 &  & \multicolumn{5}{c}{True Model (CO)} & \multicolumn{1}{c}{} & \multicolumn{5}{c}{Misspecified Model (PE)} & \multicolumn{1}{c}{} & \multicolumn{5}{c}{Misspecified Model (ED)} \\ \cline{3-7} \cline{9-13} \cline{15-19} 
CO & $F_1(70; \mbox{TVC = 0, G = 0})$ & 13.41 & 0.02 & 1.44 & 1.43 & 0.95 &  & 13.41 & -0.29 & 1.35 & 1.45 & 0.94 &  & 13.41 & 0.55 & 1.50 & 1.43 & 0.92 \\
 & $F_1(70; \mbox{TVC = 1, G = 0})$ & 16.41 & -0.04 & 2.02 & 2.03 & 0.94 &  & 16.41 & 1.42 & 2.17 & 2.13 & 0.91 &  & 16.41 & -1.52 & 1.67 & 1.58 & 0.80 \\
 & $F_1(70; \mbox{TVC = 0, G = 1})$ & 58.82 & 0.25 & 3.15 & 3.10 & 0.94 &  & 58.82 & -0.23 & 3.19 & 3.22 & 0.93 &  & 58.82 & 1.01 & 3.16 & 3.06 & 0.92 \\
 & $F_1(70; \mbox{TVC = 1, G = 1})$ & 64.90 & 0.06 & 3.80 & 3.86 & 0.95 &  & 64.90 & 2.70 & 4.10 & 3.82 & 0.85 &  & 64.90 & -3.31 & 3.46 & 3.25 & 0.81 \\
 &  &  &  &  &  &  &  &  &  &  &  &  &  &  &  &  &  &  \\
 & $F_2(70; \mbox{TVC = 0, G = 0})$ & 5.53 & 0.05 & 0.88 & 0.93 & 0.95 &  & 5.53 & 0.08 & 1.02 & 0.96 & 0.94 &  & 5.53 & -0.09 & 0.89 & 0.92 & 0.95 \\
 & $F_2(70; \mbox{TVC = 1, G = 0})$ & 5.39 & 0.04 & 0.85 & 0.91 & 0.95 &  & 5.39 & 0.03 & 0.99 & 0.93 & 0.94 &  & 5.39 & -0.03 & 0.88 & 0.91 & 0.95 \\
 & $F_2(70; \mbox{TVC = 0, G = 1})$ & 14.61 & -0.02 & 1.38 & 1.41 & 0.94 &  & 14.61 & 0.31 & 1.46 & 1.48 & 0.95 &  & 14.61 & -0.17 & 1.39 & 1.39 & 0.93 \\
 & $F_2(70; \mbox{TVC = 1, G = 1})$ & 12.77 & -0.02 & 1.31 & 1.32 & 0.95 &  & 12.77 & -0.16 & 1.35 & 1.36 & 0.93 &  & 12.77 & 0.57 & 1.34 & 1.35 & 0.93 \\ \hline
\end{tabular}}
\end{adjustbox}
\bigskip
\end{table}

\newpage
  
 \begin{figure}
\centering
\includegraphics[height=0.35\textheight, width=1\textwidth]{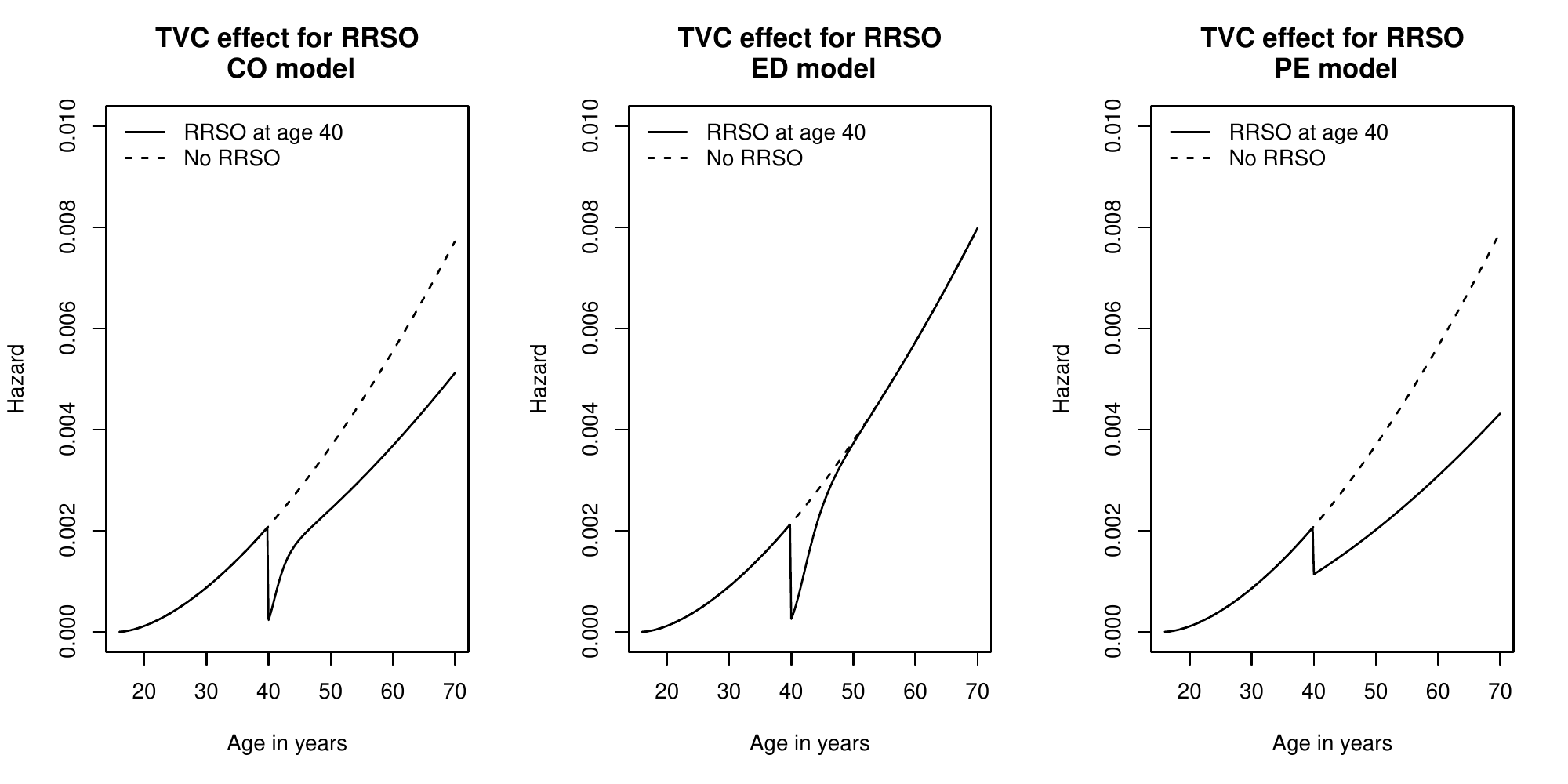}
\caption{Hazard functions estimated under the different TVC models in the {\it BRCA1} families from the BCFR.} \label{figure_hazard}
\end{figure}

\begin{figure}
\centering
\includegraphics[height=0.6\textheight, width=1\textwidth]{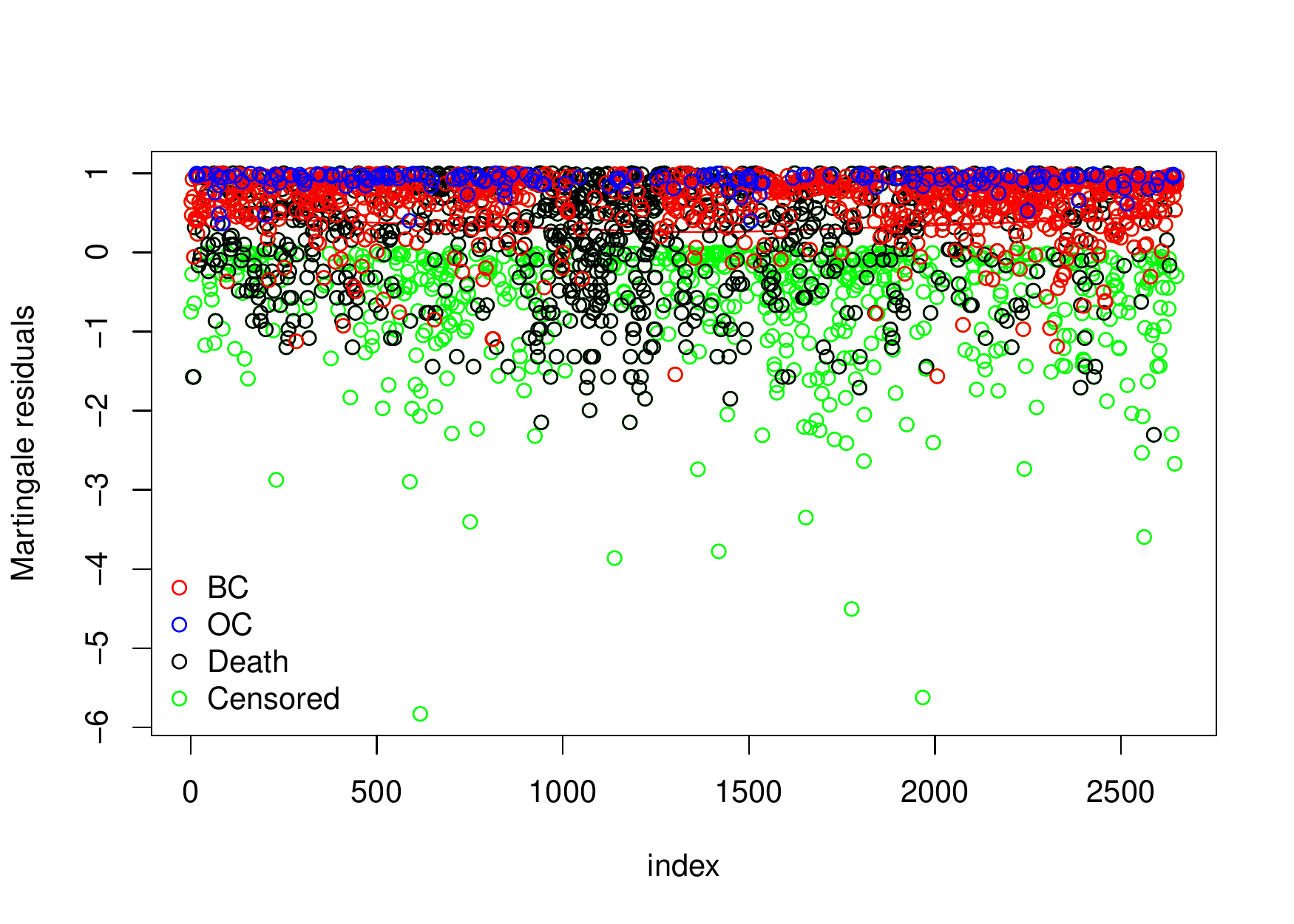} 
\caption{Martingale residuals at individual level for our best TVC model}\label{fig.mrind}
\end{figure}

\begin{figure}
\centering
\includegraphics[height=0.6\textheight, width=1\textwidth]{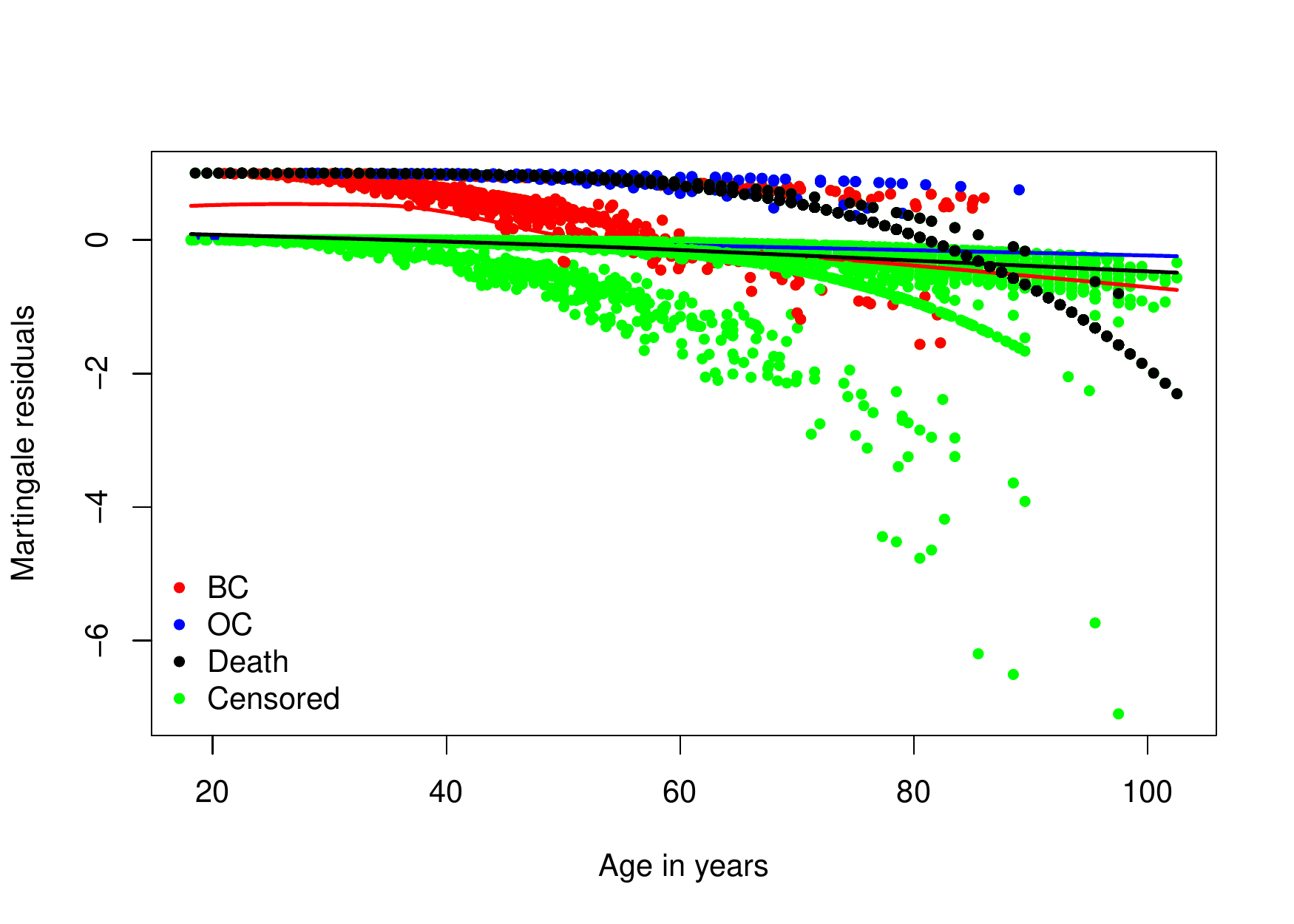} 
\caption{Martingale residuals at individual level against age in years for our best TVC model}\label{fig.mrind2}
\end{figure}

\begin{figure}
\centering
\includegraphics[height=0.6\textheight, width=1\textwidth]{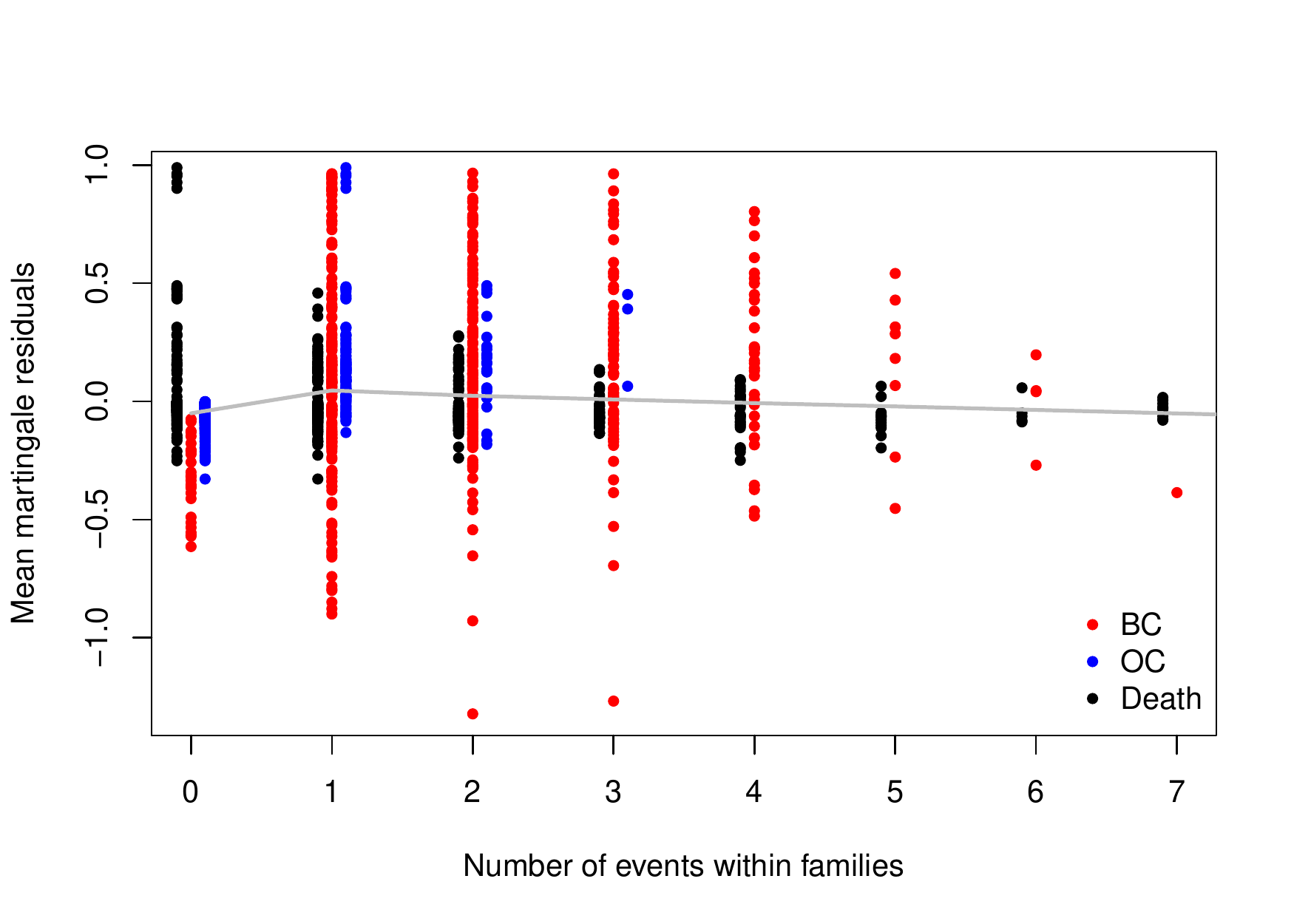} 
\caption{Martingale residuals at family level against the number of events within families for our best TVC model}\label{fig.mrfam}
\end{figure}

\end{document}